\documentclass[fleqn,10pt]{wlscirep}
\usepackage[utf8]{inputenc}
\usepackage[T1]{fontenc}
\usepackage{soul}
\newcommand{\BiSe}{Bi$_2$Se$_3$}
\newcommand{\SiO}{SiO$_2$}

\title{Quantitative infrared near-field imaging of suspended topological insulator nanostructures}

\author[1,2,$\dagger\dagger$]{C. Lupo}
\author[3]{J. Andzane}
\author[4,5]{D. Montemurro}
\author[4]{T. Bauch}
\author[4]{F. Lombardi}
\author[2]{C. Weber}
\author[1,$\dagger$]{I. Rungger}
\author[1,*]{S. E. de Graaf}
\affil[1]{National Physical Laboratory, Hampton Road, TW11 0LW, Teddington, United Kingdom}
\affil[2]{King's College London, Theory and Simulation of Condensed Matter, WC2R 2LS, London, UK}
\affil[3]{Institute of Chemical Physics, University of Latvia, LV-1586, Riga, Latvia}
\affil[4]{Department of microtechnology and nanoscience, Chalmers University of Technology, SE-41296 Goteborg, Sweden}
\affil[5]{Department of Physics “Ettore Pancini", University of Naples Federico II, c/o Complesso di Monte Sant’Angelo, via Cintia 21, I-80126 Naples, Italy}

\affil[*]{sdg@npl.co.uk}
\affil[$\dagger$]{ivan.rungger@npl.co.uk}
\affil[$\dagger\dagger$]{carla.lupo@kcl.ac.uk}

\begin{abstract} 
The development of nanoscale solid-state devices exploiting the promising topological surface states of topological insulator materials requires careful device engineering and improved materials quality. For instance, the introduction of a substrate, device contact or the formation of oxide layers can cause unintentional doping of the material, spoiling the sought-after properties. In support of this, nanoscale imaging tools can provide useful materials information without the need for complex device fabrication. Here we study \BiSe $ $ nanoribbons suspended across multiple material stacks of \SiO $ $ and Au using infrared scattering scanning near-field optical microscopy. We validate our observations against a multilayer finite dipole model to obtain quantitative imaging of the local \BiSe $ $ properties that vary depending on the local environment. Moreover, we identify experimental signatures  that we associate with quantum well states at the \BiSe $ $ surfaces. 
Our approach opens a new direction for future engineering of nanoelectronic devices based on topological insulator materials.

\end{abstract}
\begin{document} 

\flushbottom
\maketitle

\thispagestyle{empty}

Topological insulators (TIs) are an exciting class of materials that may find applications in a wide range of electronic and quantum devices, where their unique topologically protected surface states are exploited for new functionalities \cite{Hasan_2010, He_2019}. However, synthesising materials and building practical devices remains a significant challenge. Much of this can be attributed to unintentional charge doping and defects in materials resulting in the bulk dominating the electron transport, and obscuring the topological surface contribution. In addition to intrinsic charge doping by impurities or vacancies, the position of the Fermi level will be influenced by the dielectric (substrate) material on which the TI material is located \cite{Kunakova_2019}. This issue is particularly relevant for thin TI materials, where the intrinsic bulk contribution is otherwise generally minimised by reduced dimensionality ('bulk-free') \cite{Peng_2010,Xiu_2011,Felix_2018, Polyakov_2017}, and band bending will dope device surfaces differently \cite{Kunakova_2018, Kunakova_2019}. Selecting the right materials in connection to the TI is therefore important to fully exploit topological surface states. A particularly promising route for addressing this problem is the growth of single crystalline TI nanoribbons (TINRs) \cite{Andzane_2015, kong_2010}. These can easily be mechanically transferred to any substrate for further study and device fabrication\cite{montemurro2015}. Yet, device fabrication remains challenging, and obtaining transport data on a statistically relevant set of nanoribbons is very time consuming. To this end, rapid room-temperature scanning probe characterisation tools could aid in the search and characterisation of materials. Near-field optical microscopy techniques have recently emerged as effective probes of the doping of TI materials due to intrinsic defects \cite{Hauer_2015, Lu_2018} and even as potential probes of the topological surface states \cite{Mooshammer_2018}. Near-field microscopy is also a powerful tool to map surface plasmons expected to arise from the metallic TI surface states, with many applications in sub-wavelength optical devices\cite{Venuthurumilli_2019, Yuan_2017, Lu_2019, Lingstadt_2021}. In all these cases, careful modelling of the dielectric environment of the TI material is key to quantitative imaging and extraction of material properties.

Here we combine scattering scanning near-field optical microscopy (sSNOM) with theoretical modelling using a finite-dipole model (FDM) extended to multiple dielectric layers to show that local electronic properties can be extracted in TINRs extending across multiple dielectric stacks, where different substrates can induce local variations in the TINR properties. We outline the methodology for quantitative evaluation of nanoribbon properties using sSNOM and demonstrate a route towards disentangling the impact of nanoribbon substrates on material properties by suspending the TINRs across multiple substrates.

\begin{figure}[!t]
\begin{center}
\includegraphics[width=1\columnwidth]{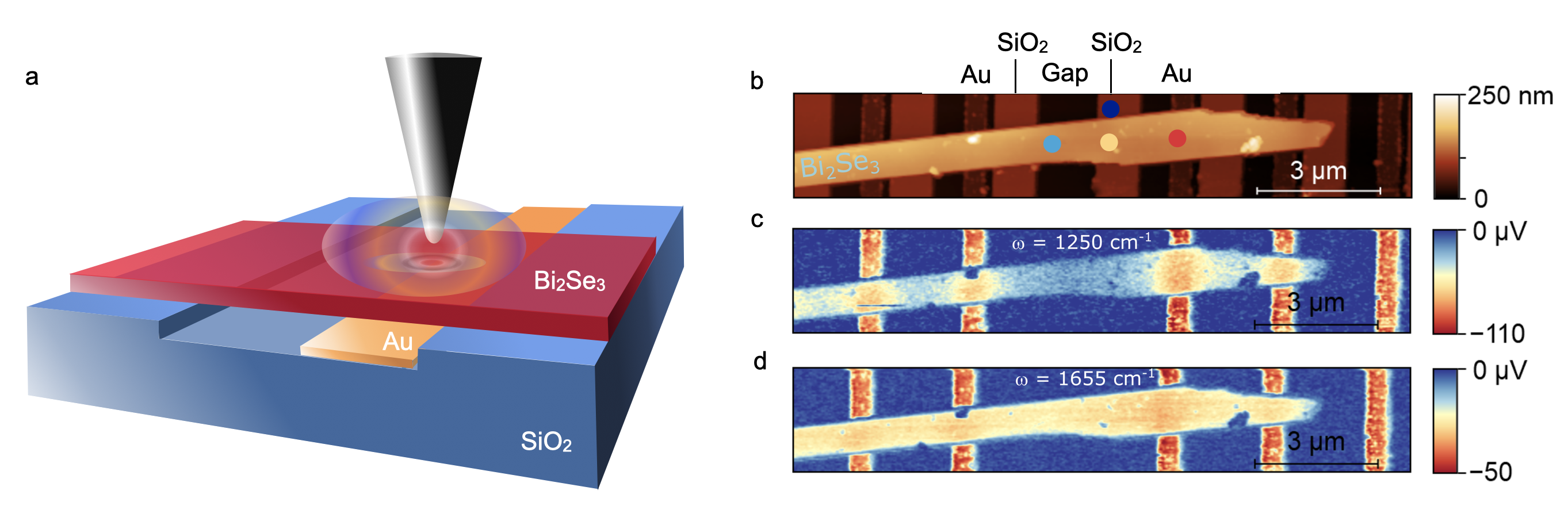}
\caption{\textbf{Set up and measurements of a \BiSe nanoribbon on a trench}. 
 \textbf{a.} Sketch of the top view of the sample in analysis and  \textbf{b.}  AFM topography showing a \BiSe $ $ nanoribbon suspended across multiple trenches in \SiO, some with Au electrodes underneath.  The four different positions, where representative measurements of the different stacks were taken, and further analysed in Figs. 2 and 3, are shown as coloured circles. \textbf{c.-d.}  sSNOM signal of the TINR for two different wavelengths, $\omega=1250$ cm$^{-1}$  and $1655$ cm$^{-1}$. The vertical stripes indicate the Au electrodes underneath the TINR. A clear variations in the signal across the TINR can be observed in  \textbf{c.}, whereas in \textbf{d.} hardly any contrast is seen. Note also the difference in colour scale.} 
\label{fig:system_set_up}
\end{center}
\end{figure}

We use \BiSe $ $ TI nanoribbons grown on a glass substrate using the catalyst-free Physical Vapour Deposition (PVD) method outlined in Ref. \cite{Andzane_2015}. The nanoribbons are mechanically transferred to a pre-patterned substrate consisting of multiple micron-sized trenches in \SiO, some with Au electrodes within. Further details can be found in the Supplemental Material (SM).  We select to study a TINR that bridges several of these trenches. The cross-section of the sample and substrate is shown in Fig. \ref{fig:system_set_up}a.

Measurements were carried out using a sSNOM operating at infrared wavelengths in combination with tapping-mode atomic force microscopy (AFM). 
The infrared near-field extends into a small volume underneath the tip, and the scattered radiation  depends on the material properties in this volume. 
To minimise the contribution from far-field effects the collected signal  is demodulated at the AFM cantilever oscillation frequency $\Omega$ (330 kHz). We denote this signal by $s_n$, where the subscript indicates the specific harmonic for demodulation. 
First, we consider surface scans  obtained at different wavelengths of the infrared radiation. 
The AFM topography in Fig. \ref{fig:system_set_up}b shows an 85 nm thick \BiSe$ $ TINR suspended across trenches in \SiO, some with Au wires inside (note that there is always a small air gap between Au and \BiSe$ $).
Fig. \ref{fig:system_set_up}c-d shows the corresponding $s_3$ sSNOM data for two selected wavelengths $\omega$. In both cases, the Au is visible as vertical stripes with strong sSNOM response, and the \SiO$ $  substrate shows almost negligible response despite significant topography variations across trenches. This indicates that the \SiO $  $ is thick enough and for theoretical modelling purposes can be taken as the bottom-most (semi-infinite) layer.   
On the TINR significant contrast variations are observed when the \BiSe $ $ is atop the Au stripes compared to a reduced signal when atop the \SiO $ $ for $\omega=1250$ cm$^{-1}$ (Fig. \ref{fig:system_set_up}c). Similarly, there is a significant difference when the TINR is suspended over the empty trenches. However, at $\omega=1655$ cm$^{-1}$ the \BiSe $ $ appears almost homogeneous (Fig. \ref{fig:system_set_up}d), with only a slightly stronger signal directly atop Au. Here only significant signal variations occurs in correspondence with the small dirt particles that causes the tip to be lifted from the \BiSe$ $ surface, resulting in a reduced signal.
These initial observations suggests that the measurement obtained by sSNOM is not exclusively dominated by the metallic gold wires, but it is sensitive to the different substrates along the nanoribbon, which may induce local variations in its properties. To provide a quantitative evaluation of the phenomena stemming from the multilayered sample, we use detailed modelling of the sSNOM response.

\begin{figure}[!t]
\begin{center}
\includegraphics[width=1\columnwidth]{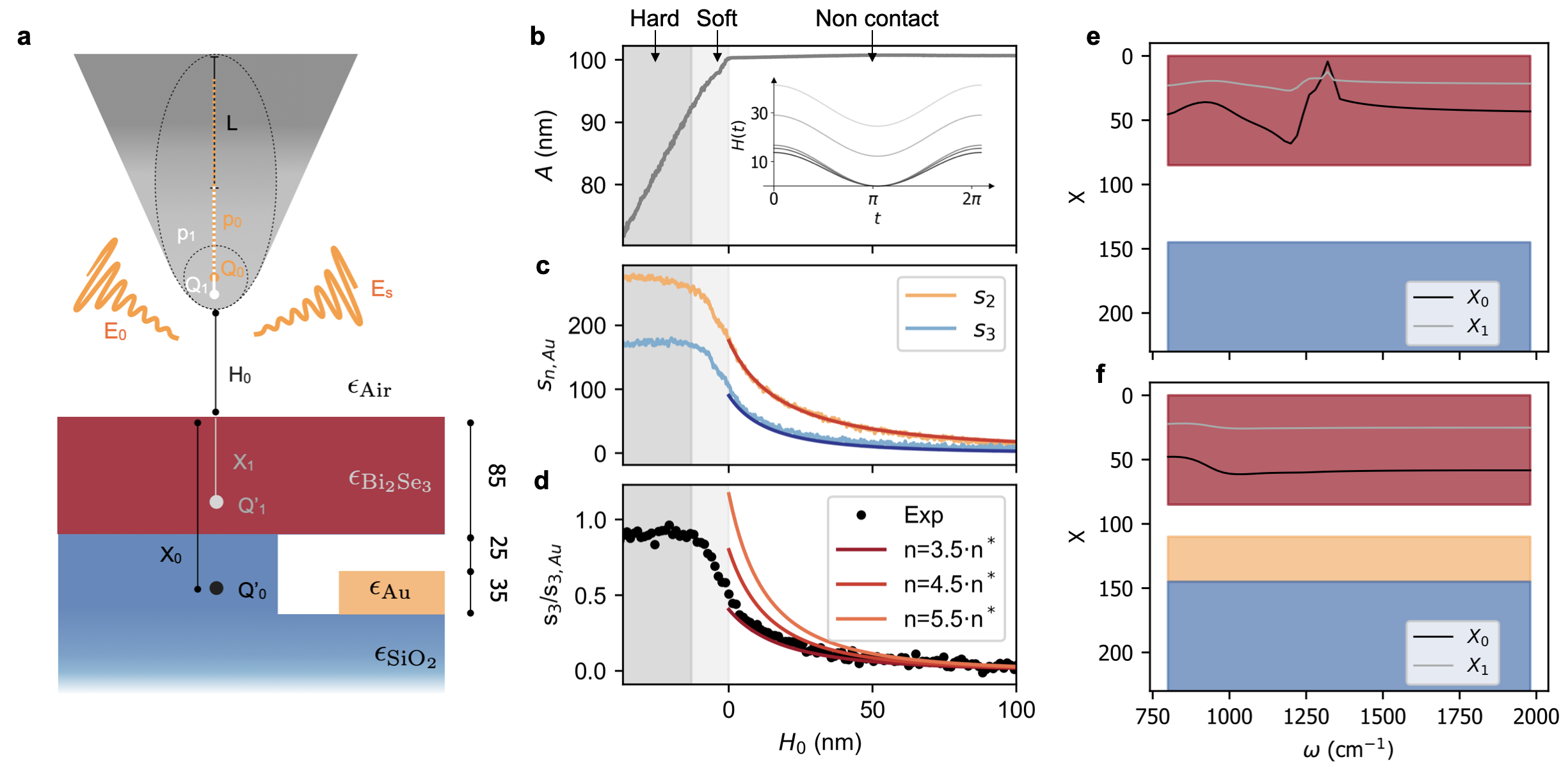}
\caption{\textbf{Validation of the MLFDM for TINR}. 
 \textbf{a.} Sketch of the tip geometry assumed in the theoretical model and the cross section of the sample in analysis.   \textbf{b.}  Re-scaled value of the measured tip oscillation amplitude as a function of the tip-sample distance (see Sec. 3 in SM for further details on $A$). When the minimum distance approaches zero, the tip enters in a \textit{soft} tapping mode where the amplitude of the oscillation is damped and the MLFDM is only approximately valid. The inset outlines the model used for the oscillating AFM tip and highlights the damping of the oscillation amplitude when the tip is in contact with the sample. \textbf{c.} Approach curves measured on Au for fixed $\omega= 950$ cm$^{-1}$ and fitted with the MLFDM model (solid line, dark shade) yielding tip parameters $R = 50$ nm, $L= 600$ nm, $g = 0.97$.  \textbf{d.} Approach curves measured on \BiSe $ $ on top of \SiO $ $ compared with model prediction obtained via the MLFDM. The results are shown for different carrier concentration, where $n^*=10^{19}$ cm$^{-3}$. Frequency dependent behaviour of the position of the image charges   \textbf{e.-f.} (in nm) in the sample.}  
\label{fig:fdm_intro}
\end{center}
\end{figure}
We model our system with a multilayer finite dipole model (MLFDM), first outlined in \cite{Cvitkovic_2007} and later discussed also in \cite{Hauer_2012} and \cite{Govyadinov_2013}. The geometry assumed in this model is shown in Fig. \ref{fig:fdm_intro}a. 
The tip is modelled by an elongated spheroid with length $2L$ and tip apex radius $R$. At the core of this model lies solving the image-charge problem as the illuminating IR radiation of amplitude $E_0$ excites an effective charge dipole, which is formed between tip and sample. 
Due to the dipoles formed by the charge-image charge pairs shown in Fig. \ref{fig:fdm_intro}a, the tip acquires an effective polarisability, which reads
\begin{equation}
 \alpha_{\text{eff}}(t)\propto 1+\frac{1}{2}\left(\frac{f_0(H(t),X_0)\beta_0}{1-f_1(H(t),X_1)\beta_1}\right) , \label{eq:alpha_def_main}  
\end{equation}
where  $X_{0/1}$ are the effective point charge positions and  $\beta_{0/1}$ are their relative quasi-static electric coefficients. $f_i(H)$ is a function of the tip geometry and $H(t)=A(1+\cos(\Omega t))+H_0$ is the tip sample distance with a static ($H_0$) and time-varying part with amplitude $A$ (see SM for further details).
Evaluating $\alpha_\text{eff}$ is key in the numerical modelling, as the scattered near-field signal $E_s$  is proportional to $\alpha_\text{eff}$. The signal is demodulated to minimise contributions from far-field effects and hence the quantity of interest is the complex-valued   $\eta_n=s_n e^{i\phi_n}=\hat{F}_n[\alpha_{\mathrm{eff}}(t)]$, arising from the Fourier transform of $\alpha_\mathrm{eff}(t)$. In what follows we refer to the measured amplitude of the near-field response $s_n$ as the detected signal at the $n^\mathrm{th}$ harmonic of $\Omega$. In general, detection at higher harmonics reduces the contribution from far-field effects. See the SM for a complete description of our model which takes into account up to 7 dielectric layers underneath the tip, sufficient to model \BiSe $ $ bulk and surface states atop 3-layer substrates, as we have here in the case of \BiSe/Air/Au/\SiO.

To calibrate the response from the tip we measure a series of approach-curves and record the detected tip oscillation amplitude $A$ (Fig. \ref{fig:fdm_intro}b) together with the near-field response (Fig. \ref{fig:fdm_intro}c) on Au. 
 We can distinguish three different regimes as function of $H_0$: \emph{(i)} \textit{non contact} regime, characterised by an $A$ independent of $H_0$.
 The point at which  $A$ starts to decrease establishes the reference height, and sets the transition into what we here call the \emph{(ii)} \textit{soft} contact regime. 
Here the tip motion is damped due to its proximity with the sample, resulting in a modification of the harmonic content of the near-field response. 
Even closer to the sample the tip motion is further perturbed, which affects the near-field harmonics generation further, resulting in a saturated near-field response. We refer to the latter as the \emph{iii)} \textit{hard} tapping regime. 
Since the FDM model is defined under the assumption of a smooth sinusoidal oscillation of the tip-sample separation, benchmark between theory and measurements is exact for near-field values recorded in the \textit{non-contact} regime, and  valid at an approximate level in the \textit{soft}-contact regime.

The geometrical properties of the tip  ($L$,  $R$, and a dimensionless parameter $g$), are then obtained through the fit of the model to the approach curves of a reference material (Au), with known optical properties. The approach curves in Fig. \ref{fig:fdm_intro}c show the behaviour of the second and the third harmonic of the near-field signal as a function of the tip-sample distance atop Au. We use Eq.\ref{eq:alpha_def_main}, with known dielectric function for Au \cite{Olmon_2012}, and we find that $ R=50$ nm, $L=600$ nm, and $g=0.97$ gives good agreement across all measured wavelengths and harmonics. 
After calibrating the tip parameters on Au, the model is further validated against the response on \SiO $ $ (see SM for further analysis).

We then proceed to extract the unknown dielectric function of the TINR located directly atop \SiO. To describe the optical response of the \BiSe$ $ we use the Drude model, whose only free parameters are inverse scattering time $\gamma_D$ and the plasma frequency $\omega_D=\sqrt{n e^2/\epsilon_0 m^*}$, where $n$ is the carrier concentration and $m^* = 0.14 m_e$ is the effective mass of bulk carriers in \BiSe \cite{orlita_2015}.  Across wavelengths a good fit is obtained with $\gamma=200$ cm$^{-1}$ and  $\omega_D = 928$ cm$^{-1}$, corresponding to $n = 3.5\cdot 10^{19}$ cm$^{-3}$. Fig. \ref{fig:fdm_intro}d shows an example of such a fit for varying $n$ at $\omega= 950$ cm$^{-1}$. This value for $n$ is in good agreement with what has been found in similar materials \cite{yan_2014, hong_2012}, and only slightly higher than found via  transport measurements of similar \BiSe $ $ TINRs grown using the same process \cite{Kunakova_2018, Kunakova_2019}, which is to be expected as device leads provides grounding and a reservoir for the carriers in the TINR.

To extract the \BiSe $ $ properties  we use the MLFDM, which takes into account the screening effect provided by the various dielectric layers underneath the tip.
The effective positions $X_{0,1}$ of the image charges, introduced in Eq. \ref{eq:alpha_def_main}, are a key indicator of this effect. 
As an example we consider two different cases: \BiSe/Air/\SiO $ $ (Fig. \ref{fig:fdm_intro}e) and \BiSe/Air/Au/\SiO$ $ (Fig. \ref{fig:fdm_intro}f).
For \BiSe/Air/\SiO, the image charge shows a frequency dependent behaviour which bears similarities with the known dielectric response of \SiO. If a layer of Au is introduced in the air gap between \BiSe $ $ and \SiO, the image charge position $X_{0/1}$ assumes a nearly constant behaviour as function of frequency, showing that even a thin layer of Au completely screens the effect of \SiO.

\begin{figure}[!t]
\begin{center}
\includegraphics[width=0.7\columnwidth]{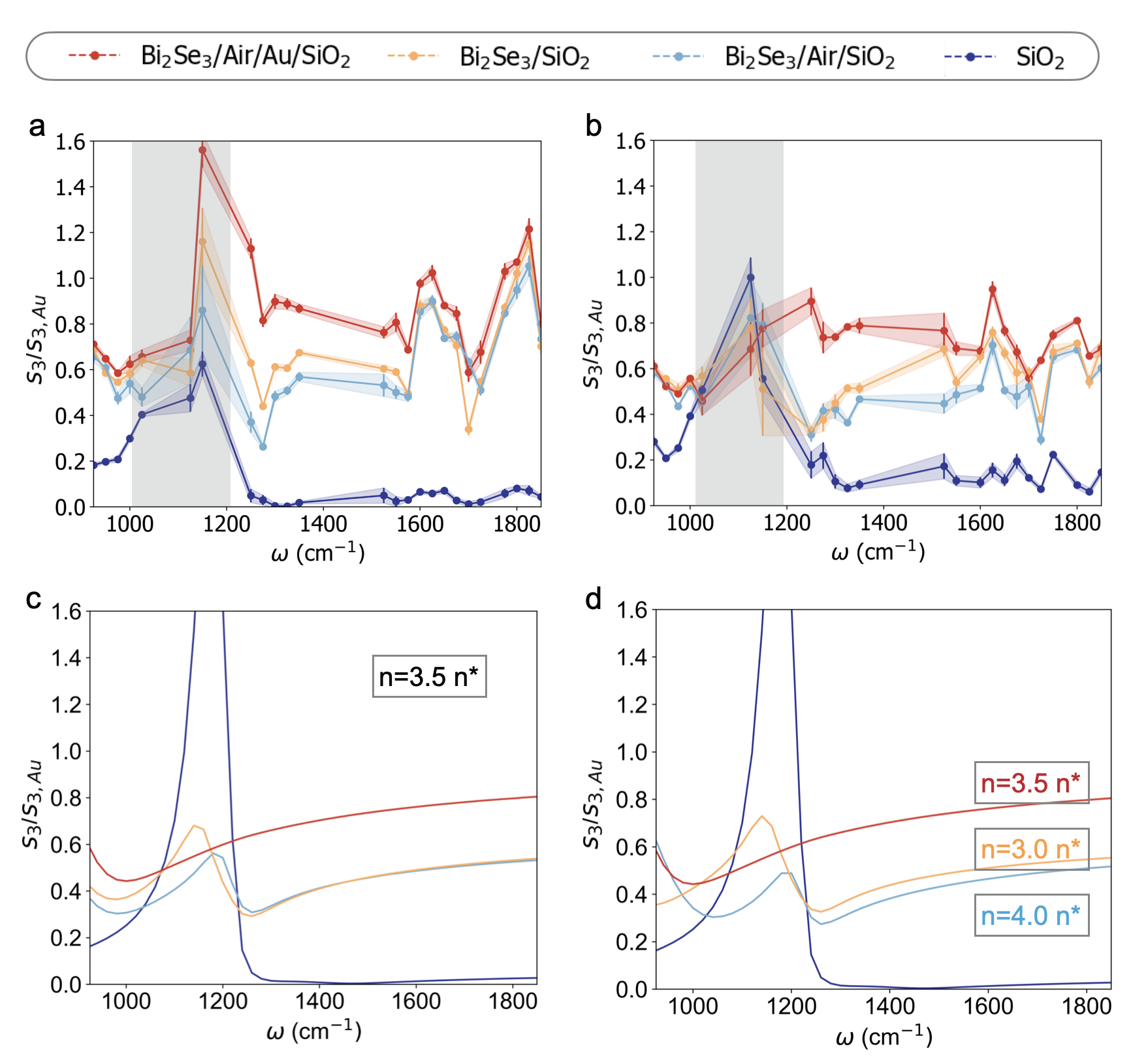}
\caption{\textbf{Probing carrier density via optical spectroscopy}. 
 Third harmonics $s_3$ of the near field signal normalised to Au and measured on \SiO  $ $ and at three different positions of the tip on the \BiSe $ $ nanoribbon.  The measurements are obtained according to two different procedures: \textbf{a.} extracted from scans with the tip in contact and \textbf{b.} from approach curves. Error bars indicate the standard deviation across the sampled pixels for each region in \textbf{a.}, while in panel \textbf{b.} they account for error stemming from the average of the approach curves in a finite small range of the tip position $H$. The grey shaded regions highlight the wavelengths at which the signal is weak therefore bearing large error bars. Panel \textbf{c.} and \textbf{d.} show the near-field signals computed with the MLFDM considering different values for the carrier concentration. In panel \textbf{c.} we characterise the \BiSe $ $ with a Drude model with constant carrier concentration $n$, while in panel \textbf{d.} $n$ changes according to the different material stacks underneath the \BiSe $ $ nanoribbon. $n^* =10^{19}$ cm$^{-3}$. The experimentally seen separation between \BiSe/\SiO $ $ and \BiSe/Air/\SiO $ $ can only be obtained by varying $n$.} 
\label{fig:frequency}
\end{center}
\end{figure}

To get an insight into the optical properties of the TINR, we analyse the response at different frequencies of the irradiating field. The spectra shown in Fig. \ref{fig:frequency}a and \ref{fig:frequency}b, are obtained using two different measurement protocols. The first approach is based on the post processing of scans. The scans are obtained in what we here refer to as the \textit{soft} tapping mode in terms of the near-field response, where the FDM is only approximately valid. Alternatively, spectra can be obtained from approach curves (Fig. \ref{fig:system_set_up}d), where the near-field signal is available within a wide range of tip-sample distances, taken at specific locations on the sample. 
In both cases we investigate the local variation of the signal in the sample and we select the measurements recorded in four different positions corresponding to different material stacks (see Fig. \ref{fig:system_set_up}b).  
Both measurements show a pronounced peak around $1100$ cm$^{-1}$,
both atop the \SiO $ $ and on top of the \BiSe. 
Across the entire frequency range considered, the strongest signal is obtained in the region where Au is below the \BiSe. A lower signal is measured when  \BiSe$ $ is in direct contact with the \SiO$ $ layer, and the weakest signal is measured when an air gap is between \BiSe $ $ and \SiO.
To investigate the  physical origin of the observed features, we compare the measurements with the MLFDM.

The model reproduces the peak at 1100 cm$^{-1}$, from the dielectric response of the \SiO. The same signature is also obtained on the \BiSe, as a consequence of the presence of the \SiO $ $ underneath the nanoribbon and a finite sampled near-field volume of size $\approx R^3$.
The MLFDM also correctly predicts that the highest near-field response in the nanoribbon is recorded when Au is underneath.  
However, we also identify some discrepancies, which we address in what follows. 

In Fig. \ref{fig:frequency}c, it can be seen that the theoretical prediction of the near-field signal is not sensitive to the presence of an air gap between the \BiSe $ $ and the \SiO $ $ for $\omega>1200$ cm$^{-1}$, yet experimental data shows a clear difference between these two stacks, both from scans (Fig. \ref{fig:frequency}a) and from approach-curve data (Fig. \ref{fig:frequency}b).
To differentiate the two predicted signals, we conducted a complete study of the role of the different parameters involved in the model.
Importantly, the observed separation between the two MLFDM curves can only be reproduced when the nanoribbon's plasma frequency is allowed to vary along its length, according to the materials below it. For a fixed $m^*$ this could only result from a varied carrier concentration $n$ (see Fig. \ref{fig:MLFDM_eta_omega_N_car_test} in SM for an extended analysis).
 This is expected to be the case, as different materials underneath may result in charge accumulation or depletion, depending on their work functions and properties of the interface \cite{Spataru_2014, Kunakova_2018} and electrostatic effects. While not expected for pristine \BiSe $ $ on \SiO $ $\cite{Kunakova_2018}, the higher $n$ observed in the case of \BiSe/Air/\SiO $ $ could arise due to the presence of the \BiSe $ $ surface oxide which modifies the work function alignment that occurs at the \BiSe/\SiO $ $ interface\cite{hong_2020}. The exact mechanism influencing the charge accumulation and depletion due to interfaces are still a matter of debate and subject to significant sample-to-sample variations, however, extensive sSNOM measurements could potentially shed more light on this issue.
We note that within nanoribbons which are transferred onto a uniform substrate (bare \SiO), we do not observe any variations, implying a constant carrier concentration. This is in contrast to previous studies in other TI materials, where local variations were found in larger structures due to defects \cite{Hauer_2012,Lu_2018}. 
Our results show that for \BiSe $ $ nanoribbons placed across multiple substrate materials the situation can be different.

These observations are valid for $\omega<1600$ cm$^{-1}$. At $\omega\sim1600$ and $\omega\sim1800$ cm$^{-1}$ we observe additional peaks in the experimental data (particularly pronounced in Fig. \ref{fig:frequency}a), which are not captured by the Drude model, but can be phenomenologically captured (see SM Fig. \ref{fig:ISS}) by adding Lorentz oscillators as additional surface layers in the MLFDM \cite{Mooshammer_2018}. 
The presence of these peaks are not expected from any of the bulk materials involved in this system. They are also weakly seen atop \SiO $ $, however, for higher harmonics, where far-field effects are suppressed, these peaks disappear (see Fig. \ref{fig:eta_omega_dh_cut} and \ref{fig:eta_omega_2D_scans}). This implies that the peaks indeed originate from the \BiSe $ $ itself. Similar features were observed in (Bi$_{0.5}$Sb$_{0.5}$)$_2$Te$_3$ \cite{Mooshammer_2018}, and are likely to originate from quantum well states (QWS) forming at the surfaces as a result of band-bending. 
Interestingly we observe two peaks of similar intensity, regardless of underlying substrate. It is therefore not likely that the two peaks originate from different QWS at top and bottom surfaces. Our simulations  show that the sSNOM should be able to detect QWS on both top and bottom surfaces with similar intensity, and we could not neglect a contribution from bottom surface despite to its larger distance from the tip. The  amorphous \BiSe $ $ oxide layer surrounding the TINR is likely to induce band-bending and result in formation of QWS on all surfaces\cite{kong_2011, hong_2020}, and the two peaks may arise from a different $m^*$ in different sub-bands.

\begin{figure}[!t]
\begin{center}
\includegraphics[width=0.7\columnwidth]{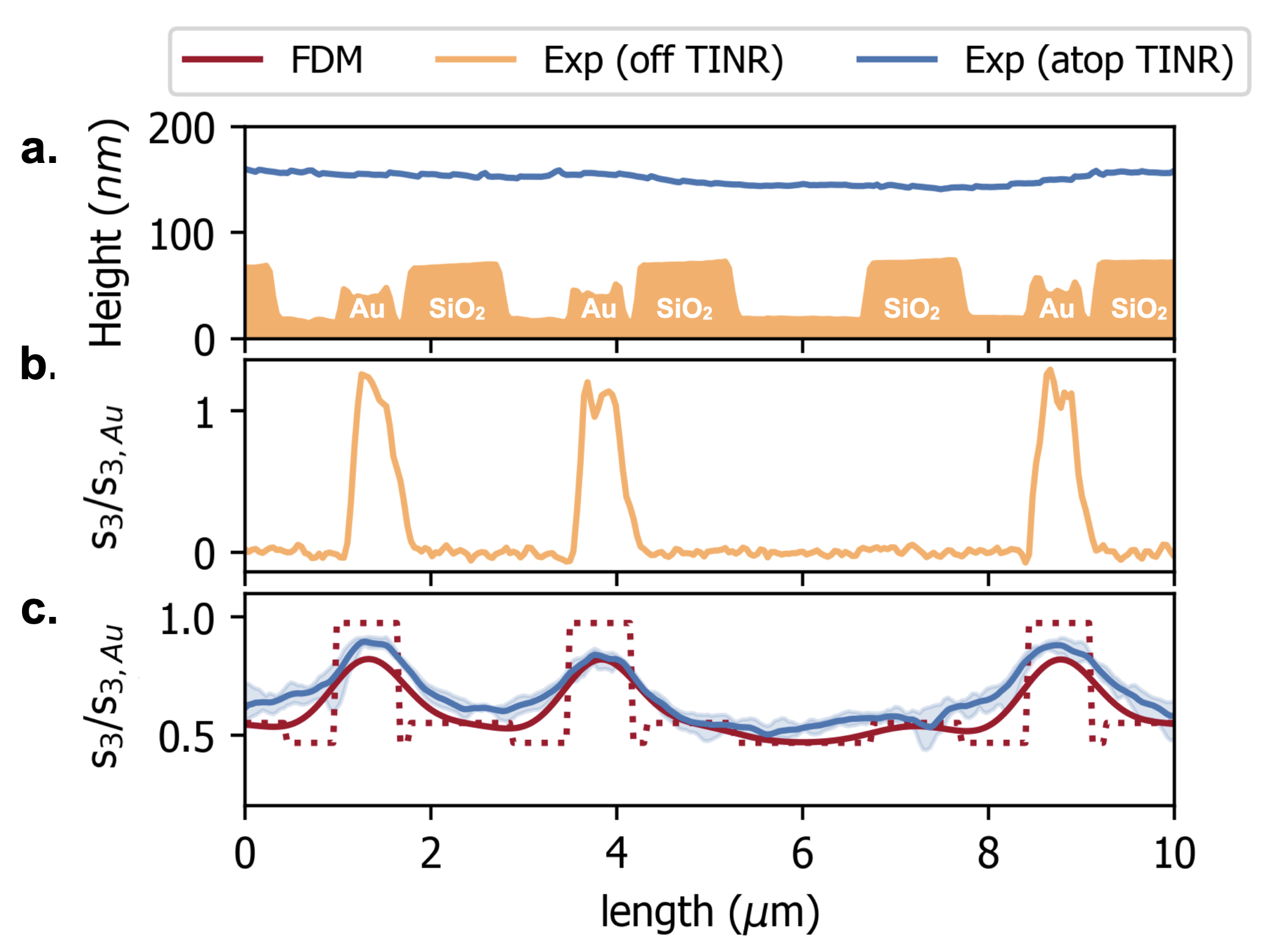}
\caption{\textbf{ 
  Scan lines and finite width effects} \textbf{a.} Profiles atop of the flat nanoribbon (blue line) and of the layers underneath measured along parallel lines on the sample. The experimental sSNOM data are taken simultaneously with the heights of the tip both off \textbf{b.} and atop \textbf{c.} the nanoribbon. In particular the middle panel reports the signal measured on the Au and \SiO across the sample.  The bottom panel shows the measurements (blue solid line) with the associated statistical errors (blue shade region). The calculated response (dotted red lines) is obtained through the MLFDM assuming a locally varying carrier concentration in three different regions as shown in Fig. \ref{fig:frequency}d. The piece-wise profile has been broadened using a Gaussian filter (solid red line) to capture finite width effects in the measurements (solid blue line).
  Measurements are reported for  $\omega=1325$ cm$^{-1}$, normalised to Au.}
\label{fig:scan_lines}
\end{center}
\end{figure}
Finally, we observe a stronger effect of the \SiO-resonance around $1100$ cm$^{-1}$ on top of the TINR (Fig. \ref{fig:frequency}a) than predicted by the model (Fig. \ref{fig:frequency}d). This is an indication that the system does not experience full screening by the TINR or Au, however, this is expected as the model describes the response of semi-infinite layers, while the measurements are taken on materials with finite dimensions. 
Another aspect of this is revealed by comparing the substrate topography (Fig. \ref{fig:scan_lines}a) with the near-field response from Au (Fig. \ref{fig:scan_lines}b), which closely follows the topography. 
On the other hand, atop the TINR, the near-field signal originates from a wider volume underneath the tip due to reduced screening of the electric field. Despite being suspended across the trenches, the topography is almost flat with corrugations uncorrelated from that of the underlying substrate (Fig. \ref{fig:scan_lines}a). The broadening observed in the near-field signal atop the TINR (Fig. \ref{fig:scan_lines}c, solid red line) is consistent with the vertical separation from the Au, as the tip is elevated by $\sim 100$ nm. To get good agreement of the step-wise theoretical solution (dashed line) we apply a Gaussian filter of width $\sigma = 340$ nm. For a vertical separation of $\sim 100$ nm, combined with the tip radius of 50 nm we would expect a broadening of $\sim 300$ nm, in good agreement.
 We also note that the differences in carrier concentrations along the TINR as deduced in Fig. \ref{fig:frequency} has been applied here, showing good agreement across the whole TINR.

In conclusion, using a finite dipole model extended to multiple layers in combination with infrared  sSNOM measurements of \BiSe $ $ nanoribbons, we studied the local variation of optical properties of the \BiSe$ $ atop different dielectric stacks, revealing local variations most likely due to changes in the carrier concentration of the \BiSe. Our results are the first application of sSNOM to infer carrier concentration locally in \BiSe$ $ nanoribbons arising from different local external interactions. 
We further identify two spectral resonances that are not captured by the independent particle model used in our approach. Corrections to the theory could entail higher order many-body effects, such as for instance Wannier-Mott excitons with large radii induced by electric field screening on the surface\cite{Wannier}, or resonances induced by surface quantum confinement due to low dimensionality\cite{Haller}. Although this extends beyond the current scope of our work, they provide promising future research directions.
The multiple substrate technique demonstrated here could thus provide direct insight into the surface properties of nanoscale TI slabs, and opens up the possibility of more detailed imaging of TINRs and devices to understand the impact of e.g. electrodes and local gates on device performance \cite{Polyakov_2017, MPlasmons_TI_2017}.

\section*{Acknowledgements}
We thank T. Vincent and A. Tzalenchuk for careful reading of the manuscript. We acknowledge support from the EU Horizon 2020 research and innovation programme (Grant Agreement No. 766714/HiTIMe) and the UK government department for Business, Energy and Industrial Strategy through the UK national quantum technologies programme. CL is supported by the EPSRC National Productivity Investment Fund (NPIF) Award and the Centre for Doctoral Training in Cross-Disciplinary Approaches to Non-Equilibrium Systems (CANES, EP/L015854/1).

\section*{Author contributions statement}
C.L. and I.R. developed the model. S.E.G. performed the sSNOM measurements and analysed the data together with C.L and I.R. J.A. synthesised the \BiSe$ $ nanoribbons, D.M. and T. B. and F.L. fabricated the substrates. C.L., S.E.G and I.R. merged theory with experimental results and wrote the manuscript. All authors discussed the results and the manuscript.

\newpage

\begin{centering} \LARGE Supplementary Material\\\vspace{3mm}
\Large Quantitative infrared near-field imaging of\\ suspended topological insulator nanostructures\\
\end{centering}

\tableofcontents

\setcounter{figure}{0}
\renewcommand{\thefigure}{S\arabic{figure}}
\setcounter{table}{0}
\renewcommand{\thetable}{S\arabic{table}}

\setcounter{equation}{0}
\renewcommand{\theequation}{S\arabic{equation}}

\section{Experimental set up}

\subsection{Sample preparation}
The sample considered here  consisted of a thermally oxidised n-doped Si wafer with a 300 nm thick \SiO $ $ layer. Trenches in \SiO $ $ were created by sputtering an additional 45 nm layer of \SiO $ $  that was subsequently etched down in selected places using reactive ion etching. In a second lithography step 20 nm thick Au electrodes were placed into some of these trenches using thermal evaporation and lift-off. The final cross-section of this sample is shown in Fig 1. Free-standing \BiSe $ $ nanoribbons were grown on a glass substrate using the catalyst-free Physical Vapour Deposition (PVD) method as outlined in \cite{Andzane_2015}. The as-grown TI nanoribbons were mechanically transferred to a pre-patterned substrate. The \BiSe $ $ nanoribbon studied here bridged several trenches and had a thickness of 80 nm, a width of 0.8-1.2 $\mu$m, and a length of several 10's of $\mu$m.
\subsection{Measurements} 
\begin{figure}
    \centering
    \includegraphics[width=1\columnwidth]{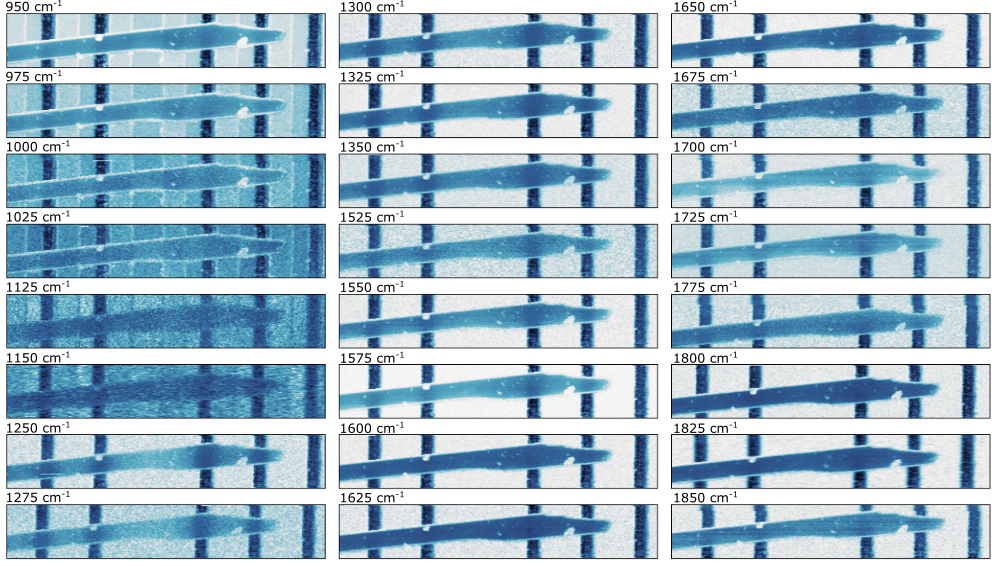}
    \caption{\textbf{Surface scans taken at different wavelengths.} Raw data showing the third harmonic near-field scans for the range of wavelengths discussed in the text. The color scale indicates the intensity of the near-field, from a low (light blue) to a high value (dark blue). The vertical size of all the scans are 2.5 $\mu$m and the horizontal size is 15 $\mu$m.}
    \label{fig:all_scans}
\end{figure}

Measurements were carried out using a scattering scanning near field optical microscope (sSNOM) operating at infrared wavelengths (Anasys NanoIR2-s). A quantum cascade laser produces the infrared light which is focused on the apex of a conductive tip of an AFM, operating in tapping mode. The detected scattered near-field signal is demodulated at harmonics of the oscillation frequency of the AFM cantilever using a lock-in amplifier.  For each studied wavelength we first collect approach curves, recording also the optical response, at different driving amplitudes of the AFM cantilever. This allows for calibration of the tip parameters by jointly fitting the approach data for different cantilever oscillation amplitudes on known materials, as outlined further below. This is repeated at all reference points (Au, SiO$_2$, different Bi$_2$Se$_3$ locations), always referencing the sSNOM interferometer position to that of Au. Finally, we obtain a surface scan, and repeat for the next wavelength. The same tip was used to collect the whole data set discussed here and shown in Fig. \ref{fig:all_scans}.

For further analysis and extraction of spectra from these scans we first align and crop each data set to account for any spatial drift in-between scans, as can be seen in some of the panels in Fig. \ref{fig:all_scans}. This is done by an edge detection algorithm applied to the AFM topography scans in both $x$- and $y$-directions. The aligned scans are then normalised to the response on Au by computing the average Au response in multiple regions across the scan where Au is present, and re-scaling the full scan data. This is done independently for each near-field harmonic. From these aligned and normalised scans we then proceed to extract spectra and cross-sections as further discussed below.

We also note that due to a limitation of the instrument only the in-phase signal from the sSNOM interferometer can be obtained, which means that we measure the scattered signal amplitude only under the assumption of a small phase shift. Thus when the underlying substrate results in a significant phase shift of the scattered light the data is no longer reliable. In the case of the measurements presented here this specifically apples to the \SiO-peak of the bare substrate at $\omega \sim 1100$ cm$^{-1}$.

\section{The Multi Layer Finite Dipole Method (MLFDM)}
\begin{figure}[!t]
\begin{center}
\includegraphics[width=0.8\columnwidth]{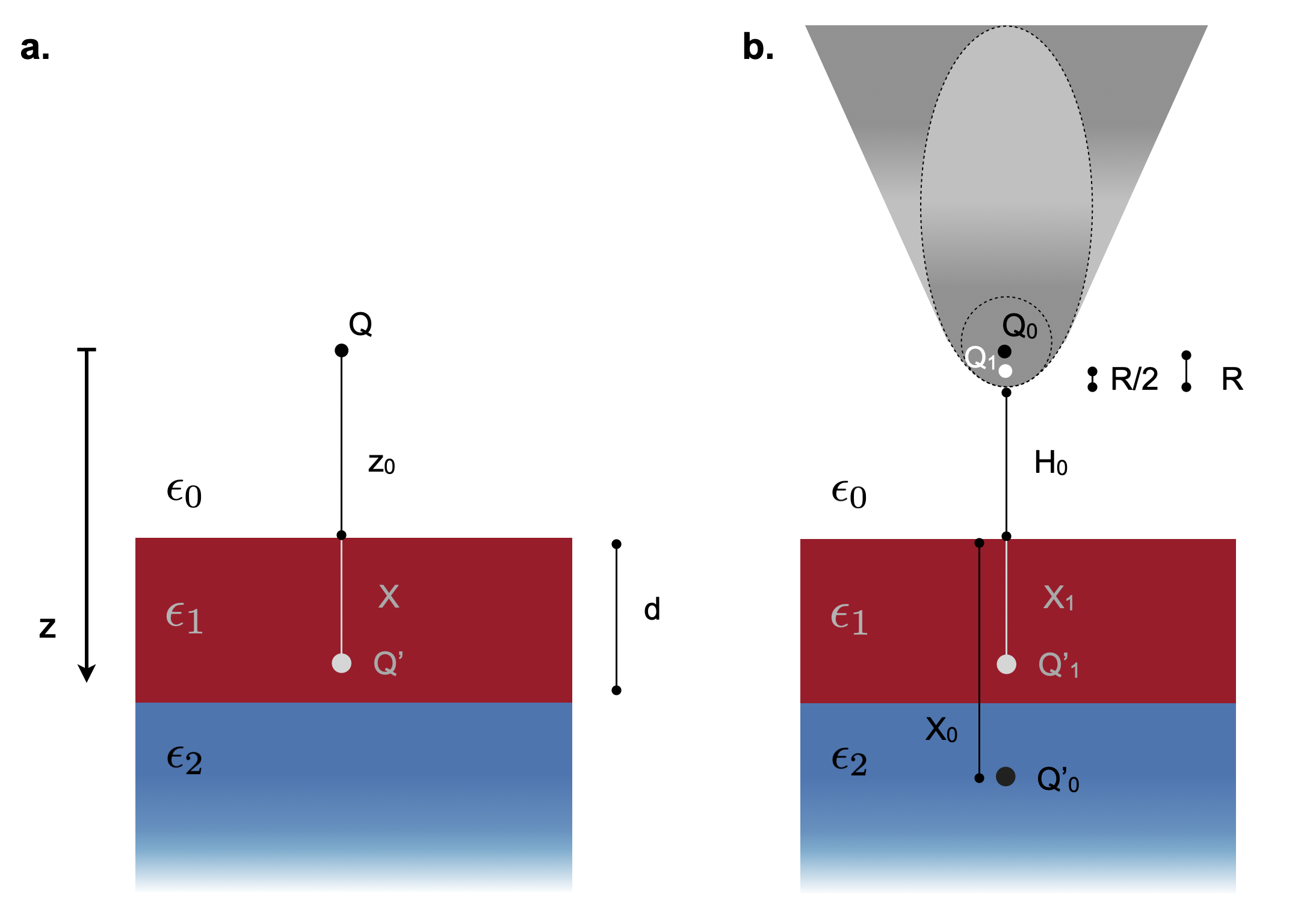}
\caption{\textbf{Derivation of the MLFDM.} Schematic picture of an AFM tip on top of a layered sample with a \BiSe $ $ nanoribbon of thickness $d$ on top of a \SiO $ $ substrate. \textbf{a.} The charge at the apex of the tip is approximated by a monopole $Q$, located at a distance $z_0$ from the sample surface. The response of the sample is approximated by a point charge $Q'$ at a distance $X$ below the sample surface. $\epsilon_0=1$ is the dielectric function of the air. The definition of the $z$ coordinate is needed to define the potential $\phi(z)$ in Eq. \ref{eq:potential_phi_Q}. \textbf{b.} Extension of the model to the multi-layers case. 
} 
\label{fig:ml_phi}
\end{center}
\end{figure}
Optical properties of the sample are obtained quantitatively using a detailed model of the tip-sample interaction for the optimal interpretation of the experimental data. In our work we make use of the so-called finite dipole model (FDM) to describe the sSNOM response. In particular we have extended upon existing formalisms  \cite{Cvitkovic_2007,Hauer_2012,Govyadinov_2013} to model our system which comprises of up to seven layers. Moreover we developed an analytical formalism which deals with complex valued dielectric functions.

Within the FDM the tip is approximated as a prolate spheroid with major axis length $2L$ and radius $R$ at the tip apex. The measured scattered field, containing the optical properties of the sample underneath the tip, is proportional to the polarisation of the tip resulting from the incident external illumination $E_0$ and the near-field interaction with the sample. Within the FDM the effective polarizability of the tip reads as \cite{Cvitkovic_2007}: 
\begin{equation}
\alpha_\text{eff}=C\left(1+\frac{1}{2}\frac{f_0(H)\beta_0}{1-f_1(H)\beta_1}\right)\label{eq:alpha_def},
\end{equation}
where $f_0(H)$ and $f_1(H)$ are functions specific to the geometry of the problem: 
\begin{equation}
f_i(H)=\left(g-\frac{(R+H+X_i)}{2L} \right)\frac{\ln(4L/(R+2H+2X_i))}{\ln(4L/R)}.
\end{equation}
Here $g$ is an empirical geometrical factor, which describes the portion of the near-field induced charge in the tip, relevant for the interaction. For typical sSNOM tip geometries, $g=0.7\pm 0.1$.
$X_{i=0,1}$ are the positions of the charges in the sample imaging the charges $Q'_0$ and $Q'_1$ in the sample, as shown in Fig. \ref{fig:ml_phi}b.

\subsection{Extension to multi-layered samples}

The extension to the multi-layer case has been obtained \cite{Hauer_2012} through the analytical solution of the boundary value problem for the electrostatic potential in ferroelectric thin films below an AFM tip \cite{Wang_2003}. In this section we provide an extension of the derivation of the multi-layer finite dipole method (MLFDM) able to describe systems consisting of up to seven layers. 

We consider a mutilayered setup as in Fig. \ref{fig:ml_phi}a. Our aim is to find how the potential above the sample is affected by the layers underneath. The potentials in the different layers are given by \cite{Wang_2003}
\begin{align}
    \phi_1(z)&=-\frac{Q}{4\pi\epsilon_0}\left[\int_0^\infty e^{-k\vert z\vert} J_0(kr) dk+\int_0^\infty A(k)e^{kz}J_0(kr)dk\right] \quad \text{for}\quad z<z_0,\\
    \phi_2(z)&=-\frac{Q}{4\pi\epsilon_0}\left[\int_0^\infty B(k) e^{-kz} J_0(kr) dk+\int_0^\infty C(k)e^{kz}J_0(kr)dk\right]\quad\text{for}\quad z_0\leq z\leq z_0+d, \\
    \phi_3(z)&=-\frac{Q}{4\pi\epsilon_0}\int_0^\infty D(k) e^{-kz} J_0(kr) dk\quad \text{for}\quad z>z_0+d.
\end{align}
The complex-valued coefficients $A(k),B(k), C(k)$ and $D(k)$ can be found by solving the boundary conditions of continuity of the potential at the interface for all values of $r\in[0,\infty)$. Thus at the boundary between the $i^{th}$ and $j^{th}$ layer, determined by the position $z=z_{ij}$, we impose:
\begin{align}
    \phi_i(z)=\phi_j(z)\vert_{z_{ij}},\\ 
    \epsilon_i\frac{d\phi_i}{dz}\vert_{z_{ij}}=\epsilon_j\frac{d\phi_j}{dz}.
\end{align}
Hence, the system of equations obtained in case of the three-layer system (see Fig. \ref{fig:ml_phi}a) reads: 
\begin{equation}
\begin{cases}
  &e^{-kz_0}+e^{kz_0}A(k)=e^{-kz_0}B(k)+e^{kz_0} C(k)\\
  &-\epsilon_0 k e^{-kz_0}+\epsilon_0 k e^{kz_0}A(k)=-\epsilon_1 k e^{-kz_0}B(k)+\epsilon_1 k e^{kz_0} C(k)\\
 &e^{-k(z_0+d)}B(k)+e^{k(z_0+d)}C(k)=e^{-k(z_0+d)}D(k)\\
  &-\epsilon_1 k e^{-k(z_0+d)}B(k)+\epsilon_1 k e^{k(z_0+d)}C(k)=-\epsilon_2 k e^{-k(z_0+d)}D(k)
\end{cases}    
\end{equation}
Solving the boundary conditions at all the interfaces in the system, we obtain
\begin{align}
    A(k)&=\left(B(k)-1\right)e^{-2kz_0}+C(k),\\
    B(k)&=\frac{2\left(1+\epsilon_1/\epsilon_2\right)}{\left(1+\epsilon_1/\epsilon_0\right)\left(1+\epsilon_1/\epsilon_2\right)-\left(1-\epsilon_1/\epsilon_0\right)\left(1-\epsilon_1/\epsilon_2\right)e^{-2kd}},\\
    C(k)&=\frac{2\left(1-\epsilon_1/\epsilon_2\right)e^{-2k(z_0+d)}}{\left(1-\epsilon_1/\epsilon_0\right)\left(1-\epsilon_1/\epsilon_2\right)e^{-2kd}-\left(1+\epsilon_1/\epsilon_0\right)\left(1+\epsilon_1/\epsilon_2\right)}.
\end{align}
In a more compact form, the potential above the sample reads  \cite{Wang_2003}
\begin{align}
\phi_Q(z)=\int_0^\infty A(k)e^{kz} J_0(kr) dk, \label{eq:potential_phi_Q} \\
A(k)=e^{-2kz_0}\frac{\beta_{01}+\beta_{12}e^{-2kd}}{1+\beta_{01}\beta_{12}e^{-2kd}}. \label{eq:A(k)_ML3}
\end{align}
Here $\beta_{ij}$ is the electrostatic reflection coefficient, which represents the key quantity for the knowledge of the optical properties of the material, and it is defined as 
\begin{equation}\label{eq:beta_ij}
\beta_{ij}(\epsilon)=\frac{\epsilon_i-\epsilon_j}{\epsilon_i+\epsilon_j},
\end{equation} 
with $\epsilon_{i}$ and $\epsilon_j$ being the dielectric functions of the two interfacing materials.

We emphasise that  the contributions from the layers underneath the tip are implicitly encoded in the function $A(k)$. Therefore, if up to $M$ layers are considered, $A(k)$ will depend on further terms, which we denote $\beta_{m-1,m}$, with $m=1\dots M$.\\
In the case of multiple layers,  $A(k)$ can be written in a compact form:
\begin{equation}
    A(k)=e^{-2kz_0}\frac{\beta_{01}+\beta_{A}e^{-2k\boldsymbol{d}}}{1+\beta_{B}e^{-2k\boldsymbol{d}}},
\end{equation}
where $\boldsymbol{d}=\sum_{i=1}^N d_i$ is the sum of the thicknesses of all the layers. The functions $\beta_{A}$ and $\beta_{B}$ changes according to the number of layers considered. Below we provide their analytical form up to 7 layers.\\

\begin{itemize}
\item MLFDM with four layers: 
    \begin{align}
    \beta_A(k)=\beta_{23}+\beta_{12}e^{2kd_2}+\beta_{01}\beta_{12}\beta_{23}e^{2kd_1}\\
    \beta_B(k)=\beta_{01}\beta_{23}+\beta_{01}\beta_{12}e^{2kd_2}+\beta_{12}\beta_{23}e^{2kd_1}
    \end{align}
\item MLFDM with five layers:
    \begin{align}
    \beta_A(k)=&\beta_{34}+\beta_{01}\beta_{12}\beta_{34}e^{2kd_1}+\beta_{23}e^{2kd_3}+\beta_{01}\beta_{12}\beta_{23}e^{2k(d_1+d_3)}+\beta_{12}e^{2k(d_2+d_3)}+\nonumber\\
    &+\beta_{01}\beta_{23}\beta_{34}e^{2k(d_1+d_2)}+\beta_{12}\beta_{23}\beta_{34}e^{2kd_2}\\
    \beta_B(k)=&\beta_{01}\beta_{34}+\beta_{12}\beta_{34}e^{2kd_1}+\beta_{01}\beta_{23}e^{2kd_3}+\beta_{12}\beta_{23}e^{2k(d_1+d_3)}+\beta_{01}\beta_{12}e^{2k(d_2+d_3)}+\nonumber\\
    &+\beta_{23}\beta_{34}e^{2k(d_1+d_2)}+\beta_{01}\beta_{12}\beta_{23}\beta_{34}e^{2kd_2}
\end{align}
\item MLFDM with six layers: 
\begin{align}
    \beta_A(k)=&\beta_{45}+\beta_{01}\beta_{12}\beta_{23}\beta_{34}\beta_{45}e^{2k(d_1+d_3)}+\beta_{01}\beta_{12}\beta_{45}e^{2kd_1}+\beta_{12}\beta_{23}\beta_{45}e^{2kd_2}+\nonumber\\
    &+\beta_{01}\beta_{23}\beta_{45}e^{2k(d_1+d_2)}+\beta_{12}\beta_{34}\beta_{45}e^{2k(d_2+d_3)}+\beta_{01}\beta_{34}\beta_{45}e^{2k(d_1+d_2+d_3)}+\beta_{34}e^{2kd_4}+\nonumber\\
    &+\beta_{01}\beta_{12}\beta_{34}e^{2k(d_1+d_4)}+\beta_{23}e^{2k(d_3+d_4)}+\beta_{01}\beta_{12}\beta_{23}e^{2k(d_1+d_3+d_4)}+\beta_{12}e^{2k(d_2+d_3+d_4)}\nonumber\\
    &+\beta_{23}\beta_{34}\beta_{45}e^{2kd_3}\\
 \beta_B(k)=&\beta_{01}\beta_{45}+\beta_{12}\beta_{23}\beta_{34}\beta_{45}e^{2k(d_1+d_3)}+\beta_{12}\beta_{45}e^{2kd_1}+\beta_{01}\beta_{12}\beta_{23}\beta_{45}e^{2kd_2}+\nonumber\\
    &+\beta_{23}\beta_{45}e^{2k(d_1+d_2)}+\beta_{01}\beta_{12}\beta_{34}\beta_{45}e^{2k(d_2+d_3)}+\beta_{34}\beta_{45}e^{2k(d_1+d_2+d_3)}+\beta_{01}\beta_{34}e^{2kd_4}+\nonumber\\
    &+\beta_{12}\beta_{34}e^{2k(d_1+d_4)}+\beta_{01}\beta_{23}e^{2k(d_3+d_4)}+\beta_{12}\beta_{23}e^{2k(d_1+d_3+d_4)}+\beta_{01}\beta_{12}e^{2k(d_2+d_3+d_4)}\nonumber\\
    &+\beta_{01}\beta_{23}\beta_{34}\beta_{45}e^{2kd_3}
\end{align}
\item MLFDM with seven layers: 
\begin{align}
    \beta_A=&\beta_{56}+\beta_{01}\beta_{12}\beta_{56}e^{2kd_1}+\beta_{12}\beta_{23}\beta_{56}e^{2kd_2}+\nonumber\\
    &+\beta_{01}\beta_{23}\beta_{56}e^{2k(d_1+d_2)}+\beta_{23}\beta_{34}\beta_{56}e^{2kd_3}+\beta_{01}\beta_{12}\beta_{23}\beta_{34}\beta_{56}e^{2k(d_1+d_3)}+\nonumber\\
    &+\beta_{01}\beta_{12}\beta_{23}\beta_{45}\beta_{56}e^{2k(d_1+d_3+d_4)}+\beta_{12}\beta_{45}\beta_{56}e^{2k(d_2+d_3+d_4)}+\beta_{01}\beta_{45}\beta_{56}e^{2k(d_1+d_2+d_3+d_4)}\nonumber\\
    &+\beta_{12}\beta_{34}\beta_{56}e^{2k(d_2+d_3)}+\beta_{01}\beta_{34}\beta_{56}e^{2k(d_1+d_2+d_3)}+\beta_{23}\beta_{45}\beta_{56}e^{2k(d_3+d_4)}\nonumber\\
    &+\beta_{34}\beta_{45}\beta_{56}e^{2kd_4}+\beta_{01}\beta_{12}\beta_{34}\beta_{45}\beta_{56}e^{2k(d_1+d_4)}+\nonumber\\
    &+\beta_{01}\beta_{23}\beta_{34}\beta_{45}\beta_{56}e^{2k(d_1+d_2+d_4)}+\beta_{45}e^{2kd_5}+\beta_{01}\beta_{12}\beta_{45}e^{2k(d_1+d_5)}\nonumber\\
    &+\beta_{12}\beta_{23}\beta_{45}e^{2k(d_2+d_5)}+\beta_{01}\beta_{23}\beta_{45}e^{2k(d_1+d_2+d_5)}+\beta_{23}\beta_{34}\beta_{45}e^{2kd_3}+\nonumber\\
    &+\beta_{34}e^{2k(d_4+d_5)}+\beta_{01}\beta_{12}\beta_{34}e^{2k(d_1+d_4+d_5)}+\beta_{12}\beta_{23}\beta_{34}e^{2k(d_2+d_4+d_5)}\nonumber\\
    &+\beta_{01}\beta_{23}\beta_{34}e^{2k(d_1+d_2+d_4+d_5)}+\beta_{23}e^{2k(d_3+d_4+d_5)}+\beta_{01}\beta_{12}\beta_{23}e^{2k(d_1+d_3+d_4+d_5)}\nonumber\\
    &+\beta_{12}e^{2k(d_2+d_3+d_4+d_5)}+\beta_{12}\beta_{23}\beta_{34}\beta_{45}\beta_{56}e^{2k(d_2+d_3+d_4)}\nonumber\\
    &+\beta_{01}\beta_{12}\beta_{23}\beta_{34}\beta_{45}e^{2k(d_1+d_3+d_5)}+\beta_{12}\beta_{34}\beta_{45}e^{2k(d_2+d_3+d_5)}+\beta_{01}\beta_{34}\beta_{45}e^{2k(d_1+d_2+d_3+d_5)}
\end{align}
\begin{align}
    \beta_B=&\beta_{01}\beta_{56}+\beta_{12}\beta_{56}e^{2kd_1}+\beta_{01}\beta_{12}\beta_{23}\beta_{56}e^{2kd_2}+\nonumber\\
    &+\beta_{23}\beta_{56}e^{2k(d_1+d_2)}+\beta_{01}\beta_{23}\beta_{34}\beta_{56}e^{2kd_3}+\beta_{12}\beta_{23}\beta_{34}\beta_{56}e^{2k(d_1+d_3)}+\nonumber\\
    &+\beta_{12}\beta_{23}\beta_{45}\beta_{56}e^{2k(d_1+d_3+d_4)}+\beta_{01}\beta_{12}\beta_{45}\beta_{56}e^{2k(d_2+d_3+d_4)}+\beta_{45}\beta_{56}e^{2k(d_1+d_2+d_3+d_4)}\nonumber\\
    &+\beta_{01}\beta_{12}\beta_{34}\beta_{56}e^{2k(d_2+d_3)}+\beta_{34}\beta_{56}e^{2k(d_1+d_2+d_3)}+\beta_{01}\beta_{23}\beta_{45}\beta_{56}e^{2k(d_3+d_4)}\nonumber\\
    &+\beta_{01}\beta_{34}\beta_{45}\beta_{56}e^{2kd_4}+\beta_{12}\beta_{34}\beta_{45}\beta_{56}e^{2k(d_1+d_4)}+\nonumber\\
    &+\beta_{23}\beta_{34}\beta_{45}\beta_{56}e^{2k(d_1+d_2+d_4)}+\beta_{01}\beta_{45}e^{2kd_5}+\beta_{12}\beta_{45}e^{2k(d_1+d_5)}\nonumber\\
    &+\beta_{01}\beta_{12}\beta_{23}\beta_{45}e^{2k(d_2+d_5)}+\beta_{23}\beta_{45}e^{2k(d_1+d_2+d_5)}+\beta_{01}\beta_{23}\beta_{34}\beta_{45}e^{2kd_3}+\nonumber\\
    &+\beta_{01}\beta_{34}e^{2k(d_4+d_5)}+\beta_{12}\beta_{34}e^{2k(d_1+d_4+d_5)}+\beta_{01}\beta_{12}\beta_{23}\beta_{34}e^{2k(d_2+d_4+d_5)}\nonumber\\
    &+\beta_{23}\beta_{34}e^{2k(d_1+d_2+d_4+d_5)}+\beta_{01}\beta_{23}e^{2k(d_3+d_4+d_5)}+\beta_{12}\beta_{23}e^{2k(d_1+d_3+d_4+d_5)}\nonumber\\
    &+\beta_{01}\beta_{12}e^{2k(d_2+d_3+d_4+d_5)}+\beta_{01}\beta_{12}\beta_{23}\beta_{34}\beta_{45}\beta_{56}e^{2k(d_2+d_3+d_4)}\nonumber\\
    &+\beta_{12}\beta_{23}\beta_{34}\beta_{45}e^{2k(d_1+d_3+d_5)}+\beta_{01}\beta_{12}\beta_{34}\beta_{45}e^{2k(d_2+d_3+d_5)}+\beta_{34}\beta_{45}e^{2k(d_1+d_2+d_3+d_5)}
\end{align}
\end{itemize}
In our calculations we use the multi-layer model to describe the different stacks in the sample. In particular, we use a 3-layer model to describe the stack Air/\BiSe/\SiO, a 4-layer model for Air/\BiSe/Air/\SiO $ $ and a 5-layer model for Air/\BiSe/Air/Au/\SiO. Moreover, a higher number of layers allows a more detailed characterisation of the TINR split up into 3 layers: a middle bulk layer, and two thinner top and bottom surface layers for an approximate model of the TI surface states. 

In the FDM formalism, the potential response above the sample is approximated by the potential $\phi_{Q'}$ of a point charge $Q'=-\beta_X Q$ at a distance $X$ under the sample surface. In particular
\begin{equation}\label{eq:phi_Q'}
\phi_{Q'}(z)=\frac{\beta_X Q}{\sqrt{\left[z-(X+z_0)\right]^2}}, 
\end{equation}
where $z-(z_0+X)$ is the distance from $Q'$ along the z direction as shown in Fig. \ref{fig:ml_phi}a. 
To determine the values of $\beta_X$ and $X$ we impose that, at $z=0$, the potential $\phi_{Q'}(z)$ (in Eq. \ref{eq:phi_Q'}) and electric field component in the $z-$direction coincide with the response  $\phi_{Q}(z)$ (in Eq. \ref{eq:potential_phi_Q}). Thus 
\begin{equation}
    \phi_{Q'}(z=0)=\phi_{Q}(z=0).
\end{equation}

Therefore:
\begin{equation}\label{eq:beta_X_MLFDM} 
    X=\frac{|\phi_{Q}(z)|}{\partial_z|\phi_{Q}(z)|}\bigg|_{z=0}-z_0 \quad \rm{and}\quad \beta=-\frac{\phi^2_Q(z)}{\partial_z \phi_Q(z)}\bigg|_{z=0}.
\end{equation}
Notice that, different from Eq.(9)-(10) in Ref. \cite{Hauer_2012}, it is crucial to consider the absolute value of the potential to assure that $X$ is defined as a real number. The phase is then encoded in the effective quasi-static reflection coefficient $\beta_X$.

Therefore, for a given multi-layered system as shown in Fig. \ref{fig:ml_phi}b, we first compute the integral in Eq. \ref{eq:potential_phi_Q}, we then calculate the $X_{0,1}$ and $\beta_{0,1}$  from Eq. \ref{eq:beta_X_MLFDM} and evaluate n$^{th}$ harmonic of the near-field signal given by Fourier transform of the effective polarizability of the tip $\alpha_{eff}(t)$ in Eq. \ref{eq:alpha_def}. 

\subsection{Analysis} 
\begin{figure}
    \centering
    \includegraphics[width=1\columnwidth]{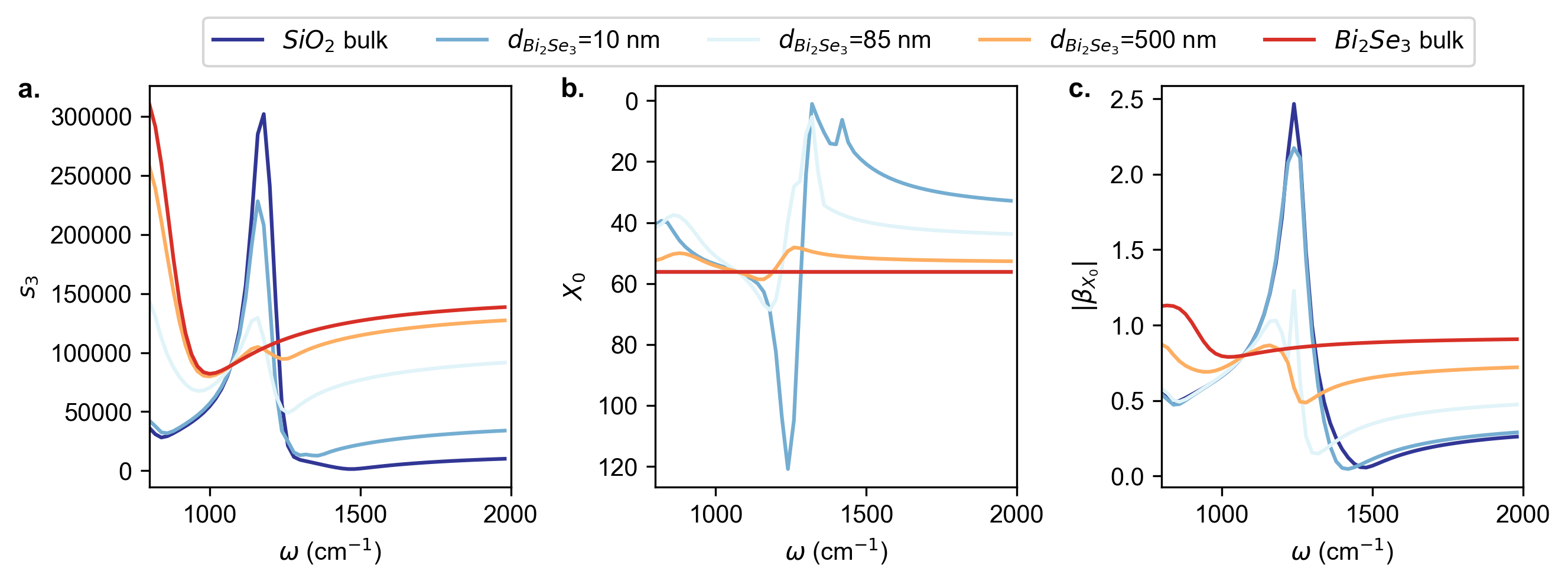}
    \caption{\textbf{MLFDM for \BiSe $ $ on \SiO $ $ substrates: dependence on the thickness of the \BiSe.} \textbf{a.} Theoretical predictions for the third harmonics of the near-field signal $s_3$ computed for a multi-layer system with a layer of \BiSe $ $ with thickness $d$ atop a substrate of \SiO. Frequency dependent behaviour of the effective position $X_0$ \textbf{b.} and relative quasi-static reflection coefficient $\beta_0$ \textbf{c.} of the image charge $Q'_0$ in the sample. The results of bulk \SiO $ $ and \BiSe $ $ are also shown for comparison. The tip parameters used for the MLFDM are $R=50$ nm, $L=600$ nm, $g=0.97$, and $A=100$ nm, while $n=3.5\cdot 10^{19}$ cm$^{-3}$ is the carrier concentration considered for \BiSe.}
    \label{fig:BiSe_SiO2_MLFDM_trends}
\end{figure}
In Fig. \ref{fig:BiSe_SiO2_MLFDM_trends} we show the theoretical prediction for the near-field signal for different thicknesses $d$ of the \BiSe. It can be seen that as $d$ decreases, the signal $s_3$ of the \BiSe $ $ converges to the \SiO $ $ response, while for $d=500$ nm, the \BiSe$ $ signal converges to the response of bulk \BiSe, meaning that the nanoribbon screens almost completely the effect of the \SiO $ $ substrate. 
This analysis emphasises that a multi-layer model is necessary to predict the optical responses of a system characterised by thin layers. 
This is further confirmed if comparing the computed signal in Fig. \ref{fig:BiSe_SiO2_MLFDM_trends}a and the measurements in Fig. 3a-b  of the main manuscript, where it can be noticed that a bulk model for \BiSe $ $ does not provide good agreement with the measured near-field values. Moreover, the observation of the \SiO $ $ fingerprints in sSNOM measurements atop the \BiSe $ $ nanoribbon shows that the thickness of the \BiSe $ $ is appropriate to classify it as a thin layer.

In Fig. \ref{fig:BiSe_SiO2_MLFDM_trends}b we study the behaviour of the effective position $X_0$ computed using Eq. \ref{eq:beta_X_MLFDM}. We observe that  $X_0$ shows frequency-dependent features for a multi-layered sample, while in case of the bulk \BiSe $ $ or \SiO $ $, $X_0$ is constant. The latter behaviour can be further confirmed considering the convergence of the Eq. \ref{eq:A(k)_ML3} to the bulk limit. This is achieved setting $d=0$ and $\epsilon_0=\epsilon_1$ (therefore $\beta_{0,1}=0$). Under this assumption, the potential above the sample becomes: 
\begin{equation}
    \phi_Q(z)=\beta_{12}\int_0^\infty e^{-k(2z_0-z)} d(k) =\frac{\beta_{12}}{2z_0-z} 
\end{equation}
Notice that the convergence of the integral above is achieved in the limit of $z<2z_0$. We then compute the effective position of the image charge $X$ and $\beta_X$. The results read:
\begin{equation}
    X=z_0\quad \beta_X=\beta_{12}
\end{equation}
which corresponds to the expected values for a sample with a single layer, where $z_0$ in Fig. \ref{fig:ml_phi} is defined as $z_0=H_0+R$ ($z_0=H_0+R/2$) when considering $Q_0$ ($Q_1$). 

In Fig. \ref{fig:BiSe_SiO2_MLFDM_trends}c we outline the behaviour of the quasi-static electric coefficient $\beta_{0}$ relative to the point charge $Q'_0$ at a position $X_0$ below the sample surface. We remind that the this is a key quantity as it encodes the optical responses of the multi-layered sample. Its behaviour resembles the one of the near-field signal (see Fig. \ref{fig:BiSe_SiO2_MLFDM_trends}a). The quenching of the peak at $\omega=1200$ cm$^{-1}$ outlines that the screening of the \SiO $ $ increases when a thicker layer of \BiSe $ $ is considered. 

These observations emphasise the important role of the multi-layer model in providing a quantitative description of the sSNOM measurements.

\section{From Model to Measurements}
In this section we outline how we fit the model to the measured approach curves.
We start by describing the obtained measurements  and their post-processing. This initial step is then  followed by the calibration of tip parameters through the fit of the measurements with the theoretical predictions.
\subsection{Analysis of the raw data}
\begin{figure}[t]
    \centering
    \includegraphics[width=1\columnwidth]{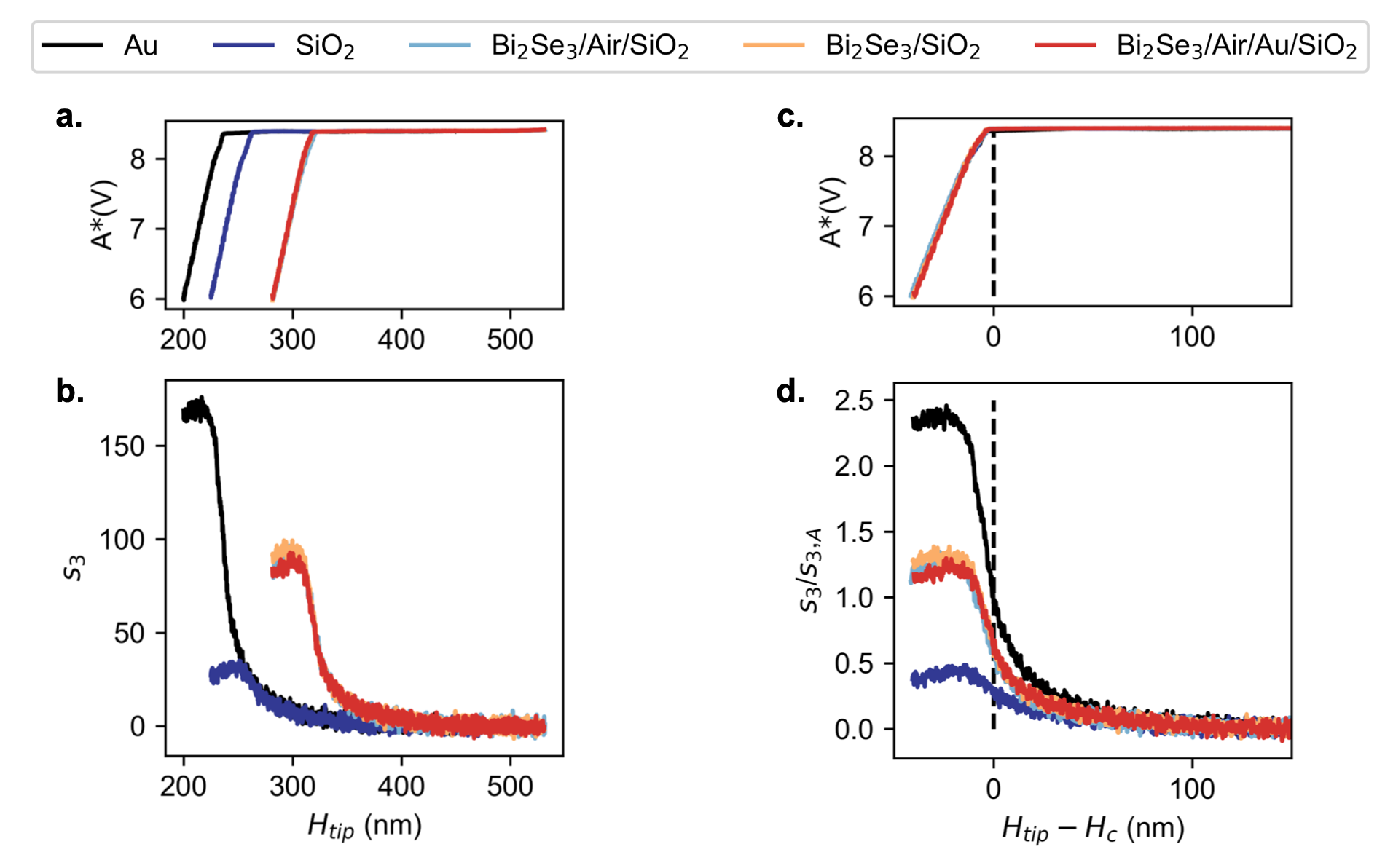}
    \caption{\textbf{Raw data (a. and b.) and post-processed data (c. and d.) of the measured approach curves.} Simultaneously recorded oscillation amplitude $A^*$ \textbf{a.} and third harmonics of the near field signal $s_3$ \textbf{b.} for different probed regions in the sample as described in Fig. 1b (in the main text), in function of the tip equilibrium position $H_\text{tip}$. Post-processed data \textbf{c.-d.} as described in the text. In particular the initial raw data shown in panel \textbf{b.} have been \emph{i)} shifted along the x-axis by the reference distance at which the tip is in contact with the sample $H_c$ \emph{ii)}  shifted along the y-axis by the value of the background signal measured from the tails of each force curve and \emph{iii)}  rescaled by the value of the near-field signal at $H_0=0$ of the reference material (Au, data set A). Note that $H_0=H_{tip}-H_c$ is the equilibrium tip-sample distance, as shown in Fig. \ref{fig:ml_phi}.}
    \label{fig:approach_curves_processing}
\end{figure}

The approach curves shown in Fig. \ref{fig:approach_curves_processing}b  outline the behaviour of the near field signal in function of the tip position. Notice that the tip is in tapping mode, with an oscillation frequency of $\sim 300$ kHz, 9 orders of magnitude slower than the IR frequency,hence it can be regarded as stationary at any instant, with equilibrium position $H_{tip}$.
The raw data, shown in Fig. \ref{fig:approach_curves_processing}b, do not report explicitly the information of the tip-sample distance $H_0$, but rather the near-field signal is recorded as function of the tip's absolute height $H_{tip}$. As a  consequence, the approach curves measured at different positions  of the sample do not have an aligned zero height value. 
To proceed with the modelling, we need to extract the reference distance of the tip in contact with the sample. To this aim we use the recorded oscillation amplitude $A^*$.

The raw data for the tip oscillation amplitude shown in Fig. \ref{fig:approach_curves_processing}a is the photodetector response (in volts) as a red laser illuminates the back of the oscillating cantilever, as in a conventional AFM. The measured $A^*$ represents the first harmonic of the oscillating laser intensity.

From Fig. \ref{fig:approach_curves_processing}a we can see that the tip is oscillating at a constant amplitude up to a critical height, below which it starts to decrease. 
We associate the inflexion point of $A^*$ to the reference position, where tip gets in contact with the sample at the point of maximal extension during the oscillations. We set this point as the boundary between \textit{non-contact} and \textit{contact} tapping mode. In the \textit{contact} tapping mode, the motion of the tip is distorted for part of the oscillation and the measured oscillation amplitude decreases. Note that for heights only slightly below this point the distortion of the oscillation from a pure sinusoidal is small. The details of the modelling of this behaviour are reported in the next section, where we discuss how to calibrate the tip parameters. In particular, to convert the data in volts $A^*$ to actual oscillation amplitudes $A$ in nanometers, the scaling factor $a=A/A^*$ is obtained via the fitting procedure discussed in the next section.  

Once the reference height described above is obtained for each curve, the curves in the data set can be shifted along the \textit{x}-axis, resulting in all being aligned with respect to the tip sample distance $H_0$, see Fig. \ref{fig:approach_curves_processing}c-d.

The next step of the post-processing involves the shift of the data along the \textbf{y}-axis by removing  the background signal, which is represented by the non-zero near-field value observed in the approach curves at long distances from the sample.  Analysing the measurements performed on the system of interest, we note that the background signal is more relevant for low harmonics ($s_2$) than higher harmonics ($s_3$) which are instead less affected by far-field effects.
We determine the value of the background signal averaging the data at the tail of the signal, in a range between 120 and 150 nm height from the sample (Fig. \ref{fig:approach_curves_processing}c-d). Hence, it is important that the measurements are performed up to sufficiently large tip-sample separation to fully capture far-field effects. At last, the signals must be re-scaled with respect to a known reference signal, which in our case is represented by the signal of Au at $H_0=0$. 
We point out that the procedure shown in Fig. \ref{fig:approach_curves_processing} is for a fixed IR frequency $\omega=950$ cm$^{-1}$. The vertical shift of the background signal and the re-scaling with respect to the reference material (Au) need to be repeated for all the different frequencies in the data set.

\subsection{Tip Parameters Calibration}\label{sec:SI_tip_calibration}

\begin{figure}
    \centering
    \includegraphics[width=0.8\columnwidth]{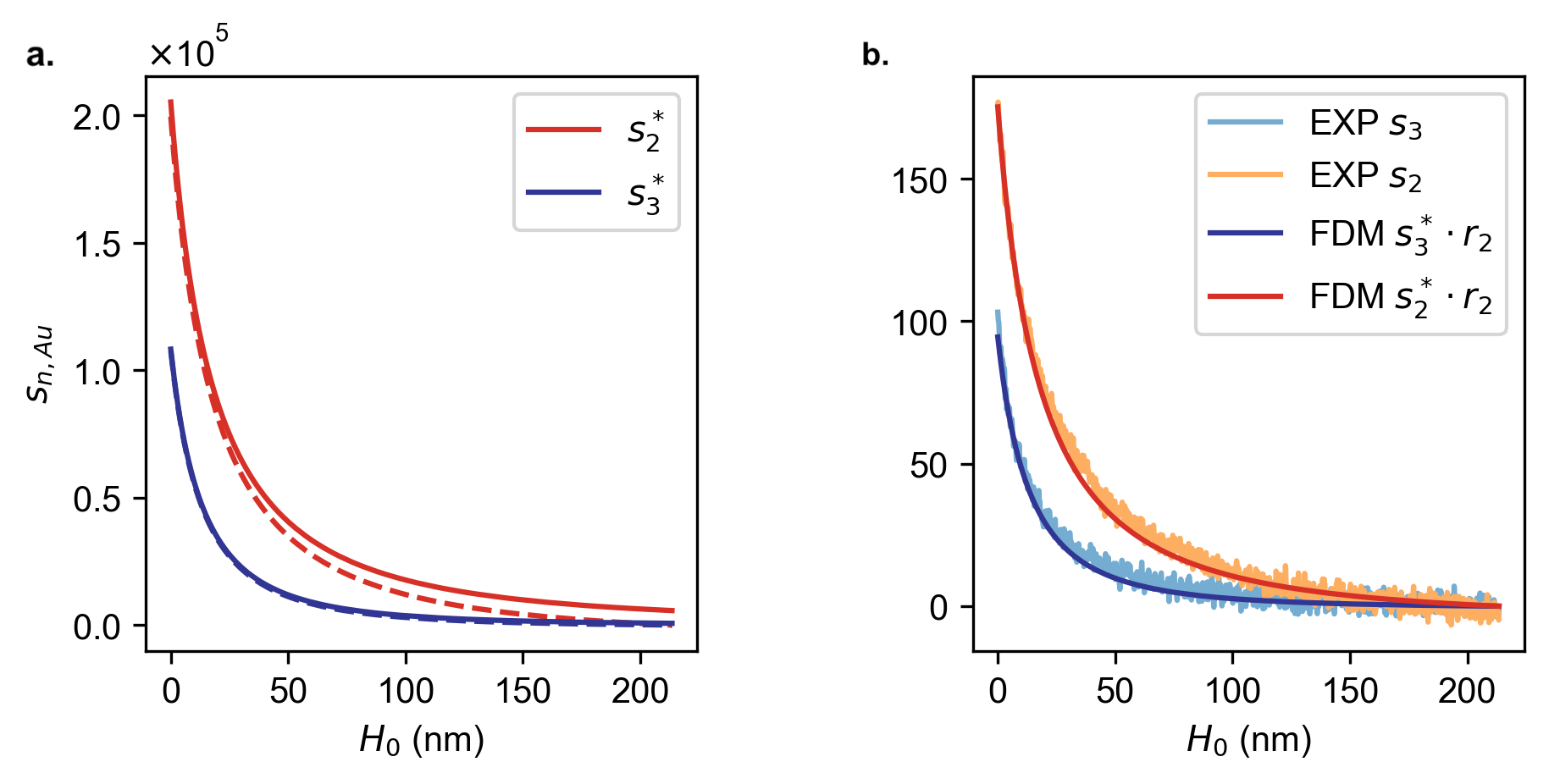}
    \caption{\textbf{The FDM approach curves.} \textbf{a.} Second $s_2$ and third $s_3$ harmonics of Au computed with the FDM model, at $\omega=950$ cm$^{-1}$ and with fixed parameters $R=50$ nm $ ,a=12,L=600$ nm $,g=0.97$. Dashed and solid lines refer respectively to the results with and without the shift of the background signal. Notice that the third harmonics has nearly zero offset due to far-field effects. This is expected as higher harmonics probe smaller volumes in the sample, therefore the signal decreases faster when the tip-sample distance increases. \textbf{b.} Final result of the processing of the theoretical approach curves. In particular the theoretical results have been rescaled by a factor $r_2$ in order to match the experimental force curves.}
    \label{fig:approach_curves_processing_FDM}
\end{figure}

\begin{figure}
    \centering
    \includegraphics[width=0.8\columnwidth]{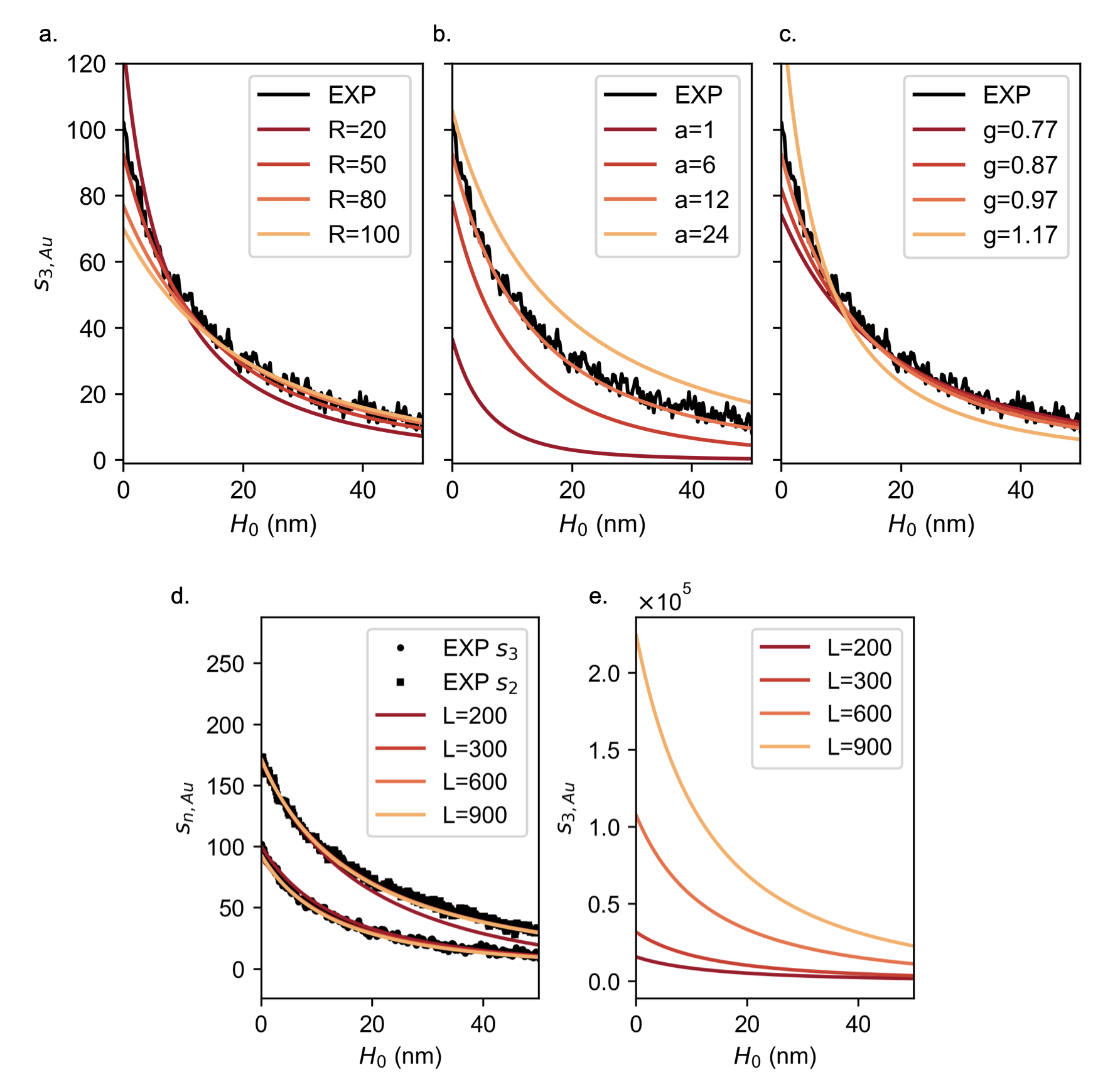}
    \caption{\textbf{Calibration of the tip geometry.} Benchmark of measured (black line) and computed approach curves for the third harmonics of the near-field signal $s_3$ of Au. The non-varying parameters in panels \textbf{a.-e.} are set to $R=50$ nm,$ L=600$ nm, $ g=0.97$, $a=12$. The frequency is fixed to $\omega=950$ cm$^{-1}$.}
    \label{fig:approach_curves_trends}
\end{figure}

The effective polarizability of the tip, introduced in Eq. \ref{eq:alpha_def}, depends on the geometry of the tip and on the optical properties of the sample considered. The unknown geometrical parameters are obtained through the fitting of the approach curves of a reference material whose optical response is well known. In our case we consider the near-field signal obtained atop Au.

\paragraph{Fitting procedure of the tip geometrical properties.} 
The FDM curves are obtained from the Fourier transform of  Eq. \ref{eq:alpha_def}, with the dielectric function of Au for a fixed wave-number $\omega$ and a set of initial guesses for $R,L,g,a$. The tip-sample distance $H_0$ and the  oscillation amplitude $A^*$ are the measured quantities, as shown in Fig. \ref{fig:approach_curves_processing}. The time varying position of the tip is described in our model by the equation
\begin{equation}
H(t) = A\left[1+\cos(\Omega t)\right]+H_0,
\end{equation}
where $A=a\cdot A^*$ is the oscillation amplitude re-scaled from Volts to nanometers.
Consistently to what is performed in the post-processing of the measured data, the near-field values obtained with the FDM model are shifted by the background signal. Secondly, the obtained theoretical approach curves for Au in  Fig. \ref{fig:approach_curves_processing_FDM}a are of a different order of magnitude with  respect to the measurements shown in  Fig. \ref{fig:approach_curves_processing}b. 
To properly match the theoretical with the experimental curve, a multiplicative factor $r_2$ is obtained considering the ratio of the two curves. 
We emphasise that this multiplicative factor is important only for getting the geometrical properties of the tip and does not play any role when re-scaling the full data set with respect to the Au reference. Furthermore, the ratio $r_2$ should not depend on the different near-field harmonics. Therefore in our procedure we compute $r_2$ from the ratio of the computed and measured second harmonics of the signal. Then, to validate this fit we re-scale the computed values for $s_3$ and compare them with the measurements. 

In Fig. \ref{fig:approach_curves_trends} we report the comparison between the measurements and the theoretical curves obtained via the FDM method, using different combinations of values of tip radius $R$, tip length  $L$, $g$-factor, and oscillation amplitude $A$. Our study reveals that the  optimal parameters are: $R=50$ nm, $g=0.97$, $A=100$ nm for the particular tip used in these experiments.

\paragraph{Tapping modes and time dependent model for the oscillating tip.}
\begin{figure}[t]
    \centering
    \includegraphics[width=1\columnwidth]{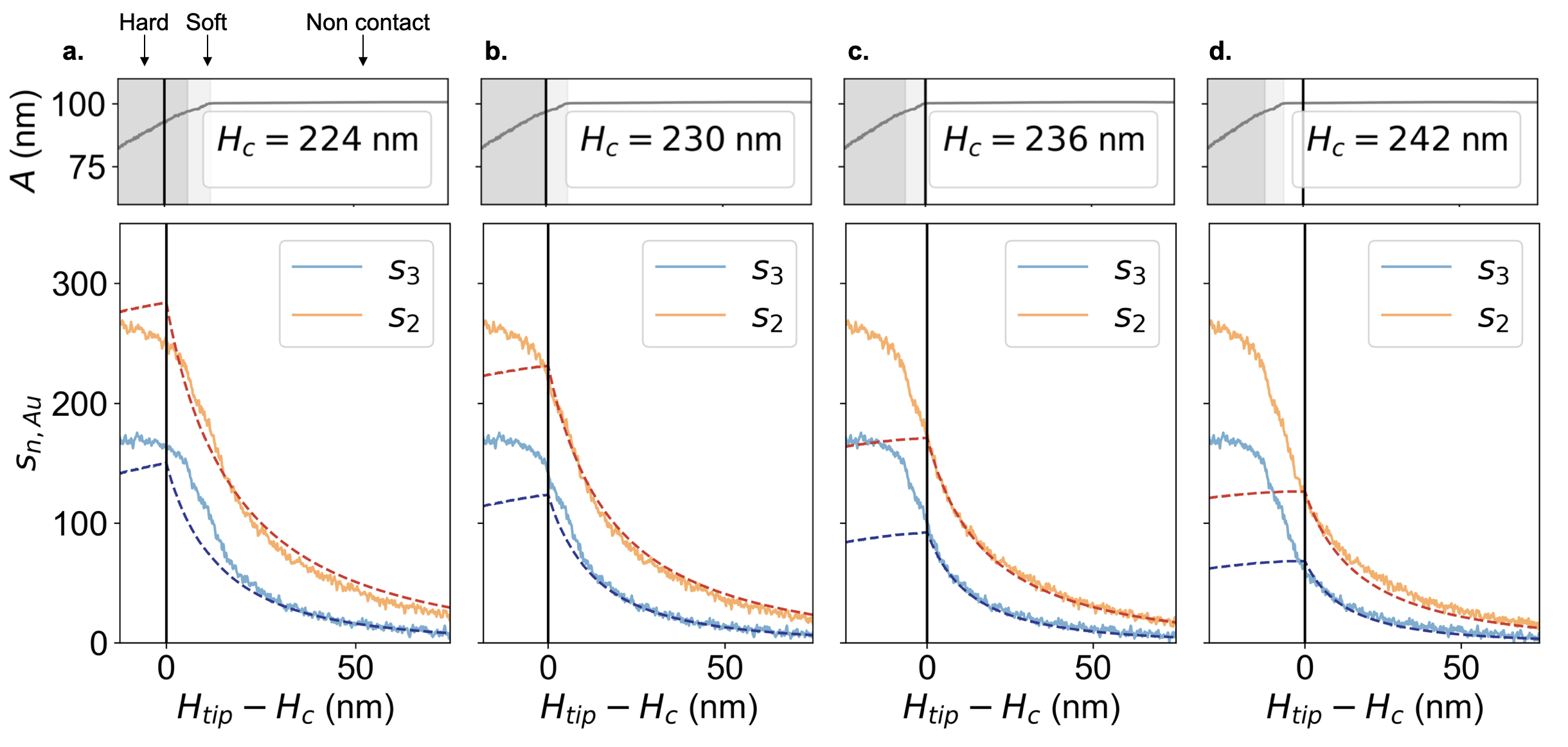}
    \caption{\textbf{FDM validity for different tapping mode regimes.} (top) Oscillation amplitude $A$ and second and third harmonics of the near-field signal obtained respectively from the measurements (solid line) and the theoretical fit (dashed lines). The different panels \textbf{a.-d.} show the result of the theoretical fit according to different selection of the distance $H_0$. The model used for $H(t)$ is described in Eq. \ref{eq:H(t)_model_complete} with $H_0=H_{tip}-H_c$ and $H_c$ being the selected reference height of the tip when in contact with the sample.}
    \label{fig:approach_curves_trends_cutH}
\end{figure}

The validity of the finite dipole method holds in the case of smooth sinusoidal behaviour of the oscillating tip. When the tip gets in contact with the sample, its motion is distorted and the measurements reveal a consequent drop of the oscillation amplitude.
Therefore, for a consistent quantitative benchmark between the theory and the measurements, we need to  distinguish the limits of the different regimes. 
As shown in Fig. \ref{fig:approach_curves_processing},  we can detect the boundary between the contact and non-contact regime from the analysis of the behaviour of the measured tip oscillation amplitude. 
In this section we analyse the validity of the model for different selection of the height $H_c$  at which we determine that the tip is in contact with the sample. The FDM curves are obtained from Eq. \ref{eq:alpha_def}, with fixed tip parameters as described above and $\omega=950$ cm$^{-1}$.
For a better description of both tapping mode regions, we extend our model for the tip as follows:
\begin{equation}
H(t) =
  \begin{cases}
    A\left[1+\cos(\Omega t)\right]+H_\text{0} & \quad \text{if } H_{0}>0 \\
    A\left[1+\cos(\Omega t)\right]    & \quad \text{otherwise }\\
  \end{cases}, \label{eq:H(t)_model_complete}
\end{equation}
where the oscillation amplitude $A$ and the distance $H_0$ are simultaneously measured.

In Fig. \ref{fig:approach_curves_trends_cutH} we show the behaviour of the oscillation amplitude and we characterise the different tapping modes. As previously stated, the inflection point $H_c$ establishes the distance at which the tip is in 'contact' with the sample and the boundary between what we here in the context of sSNOM label as the \textit{contact} and \textit{non contact} regime. A closer analysis at the contact mode region, reveals that the approach curves show the same trend as in the non contact mode, until a plateau is reached. This observation allows us to distinguish two different regimes of the contact mode, which we here label as \textit{soft} and \textit{hard} tapping regimes.

We found that the FDM model provides an exact description of the measurements performed in a \textit{non contact} regime, it shows a fair agreement if the truncation of the data is performed in the \textit{soft} contact mode, while it fails in the \textit{hard} tapping regime. Furthermore the extended Eq. \ref{eq:H(t)_model_complete} captures the observed plateau trend, albeit at incorrect values which is to be expected as Eq. \ref{eq:H(t)_model_complete} only provides a simplified picture of damped tip oscillations, and disregards any non-linearities due forces acting on the tip.
The exact description of tip dynamics able to describe the two contact regimes is left for future research. When analysing approach curves we instead focus on the data in the \textit{non contact} regime.

\subsection{Dielectric Functions}
In this section we outline the dielectric functions used to model the different materials present in the experiment.
\begin{figure}
    \centering
    \includegraphics[width=1\columnwidth]{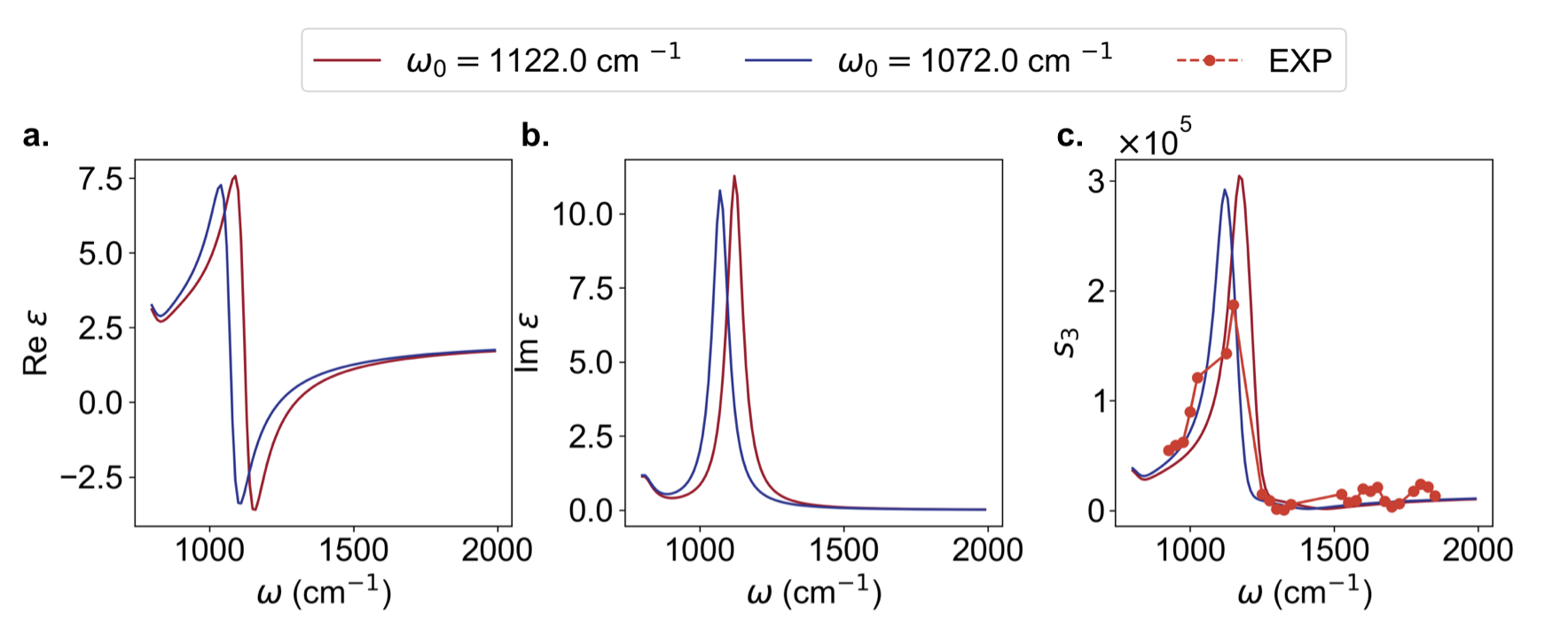}
    \caption{\textbf{Modelling $\omega_0$ for \SiO.} Real \textbf{a.} and imaginary part \textbf{b.} of the \SiO $ $ dielectric function, with different values of the plasma frequency $\omega_0$. \textbf{c.} Benchmark of the calculated third harmonics of the near-field signal $s_3$ with the measured data. In particular, the experimental data are obtained through the post-processing of the 2D scans.}
    \label{fig:SiO2_eps}
\end{figure}

\begin{figure}
    \centering
    \includegraphics[width=0.6\columnwidth]{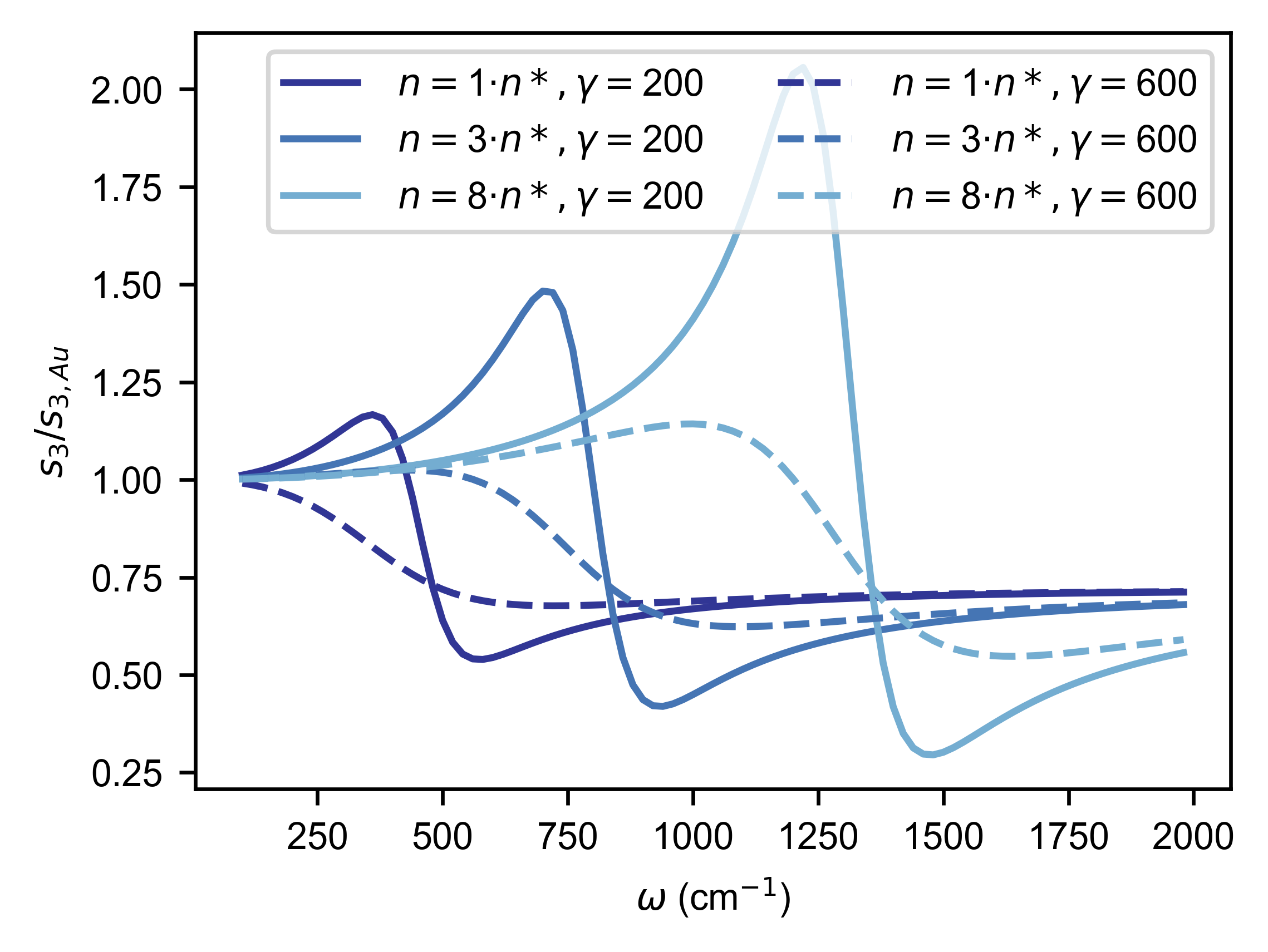}
    \caption{\textbf{Tuning of $\gamma$ for \BiSe.}  Third harmonic of the near-field signal $s_3$ of bulk \BiSe $ $ re-scaled with respect to Au. The analysis has been performed for different carrier concentrations $n$, with $n^*=10^{19}$ cm$^{-3}$. Solid and dashed lines refers respectively to different scattering rate $\gamma_D=200$ and $\gamma_D=600$ cm$^{-1}$.}
    \label{fig:BiSe_eps_gammas}
\end{figure}

\begin{table}
    \centering
    \begin{tabular}{c|c|c|c}
         $j$  &   $\omega_j$ (cm$^{-1}$)   &   $\gamma_j$ (cm$^{-1}$)    &   $f_j$ \\
         \hline
         \hline
         1  &   1122.27 &   67.2179 &   0.6752\\
         2  &   805.20  &   75.7996 &   0.0929\\
         3  &   457.61  &   44.5774 &   1.0218\\
    \end{tabular}
    \caption{Lorentz model parameters for SiO$_2$.}\label{tab:SiO2_eps}
\end{table}
\paragraph{\SiO $ $  substrate} We assume the \SiO $ $ to be the bottom-most 'bulk' layer in the FDM as its thickness (300 nm) is large enough such that anything underneath no longer contributes to the near-field response. This assumption is well justified given the much smaller tip radius and the fact that the different thicknesses of \SiO $ $ on the sample (due to the trenched regions) produces no measurable difference in the sSNOM response. The \SiO $ $ substrate is modelled in the spectral region of interest as a Lorentz oscillator:
\begin{equation}\label{eq:Lorentz}
    \epsilon(\omega)=\epsilon_\infty+\sum_{j=1}^{3}\frac{f_j\omega_j^2}{\omega_j^2-\omega^2-i\gamma_j\omega}.
\end{equation}
whose parameters are defined in Tab. \ref{tab:SiO2_eps} and $\epsilon_\infty=2.1$.

Notice that (see Fig. \ref{fig:SiO2_eps}) to properly match the experimental observations, we shifted the centre of the oscillator to $\omega_1 = 1122$ cm$^{-1}$ (instead of 1060 \cite{Lu_2018}), in agreement with what was found for thin-films of \SiO $ $ \cite{Zhang_2012}. 
\paragraph{Gold regions} The Au regions  were modelled with a Drude model with $\tau=1/\gamma=14$ fs and $h\omega_D=8.5$ eV \cite{Olmon_2012}.
\paragraph{\BiSe $ $ nanoribbons} The \BiSe $ $ nanoribbon was simulated with a Drude model, which reads  
\begin{equation}\label{eq:Drude}
\epsilon(\omega)=\epsilon_\infty\left(1-\frac{\omega_D^2}{\omega^2+i\gamma_D\omega}\right), 
\end{equation}
where $\epsilon_\infty=26$ and scattering rate $\gamma_D=200$ cm$^{-1}$.
$\omega_D=\sqrt{n e^2/\epsilon_0\epsilon_\infty m^*}/2\pi c$ with effective mass $m^*=0.14 m_e$ where $m_e$ is the free electron mass \cite{orlita_2015}. In our study the carrier concentration values which better describe the system are in a range $3-4\cdot 10^{19}$ cm$^{-3}$.

Our parameters are similar to the ones used in Ref. \cite{Lu_2018} where the \BiSe $ $ nano-crystals were modelled with a scattering rate $\gamma_D=600$  cm$^{-1}$, $m^*=0.14\cdot m_e$, $\epsilon_\infty=25$ and $n_D\approx 5-7 \cdot 10^{19}$ cm$^{-3}$.  
A better agreement with measurements is achieved in our case with $\gamma_D=200$ cm$^{-1}$. A lower scattering rate is indicative of the presence of less impurities and therefore higher quality of the synthesis. Furthermore, in Ref \cite{Wang_18} where optical properties of thin films were studied, the following parameters were found: $\omega_D=914$ cm$^{-1}$, $\epsilon_\infty=26.5$ and $\gamma_D=67$ cm$^{-1}$. The small damping parameter of the dielectric function is due to less defects arising from the epitaxial growth of the thin film.

\begin{figure}[t]
\begin{center}
\includegraphics[width=0.6\columnwidth]{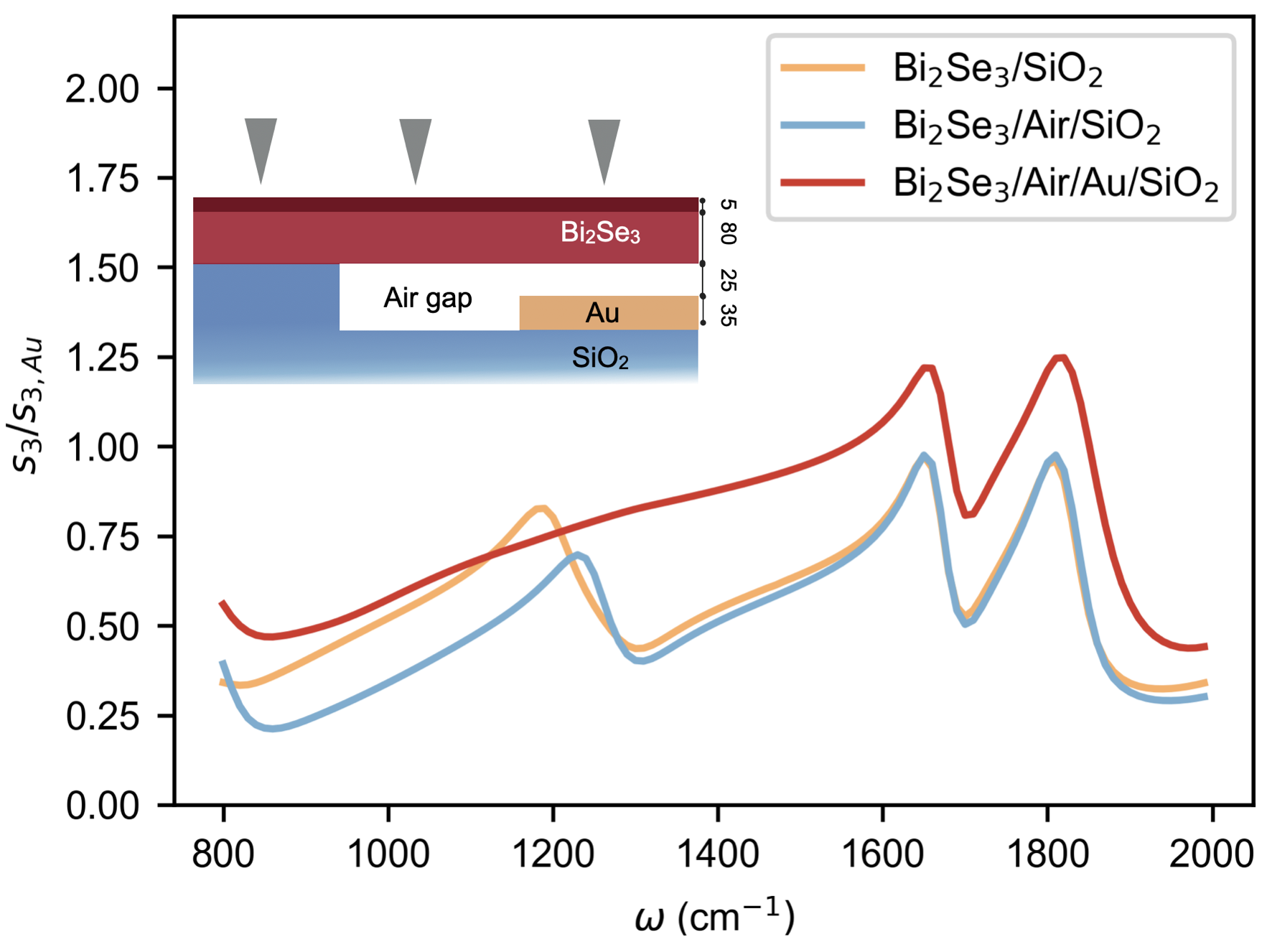}
\caption{\textbf{Additional resonances in \BiSe.}  Third harmonics $s_3$ of the near field signal normalised to Au computed with the MLFDM and assuming additional Lorentz oscillators at the nanoribbon surface, at three different positions of the tip on the \BiSe $ $ nanoribbon as outlined in the inset.}
\label{fig:ISS}
\end{center}
\end{figure}
\begin{table}
    \centering
    \begin{tabular}{c|c|c|c}
         $j$  &   $\omega_j$ (cm$^{-1}$)   &   $\gamma_j$ (cm$^{-1}$)    &   $f_j$ \\
         \hline
         \hline
         1  &   1650 &   40 &   20\\
         2  &   1800  &  50 &   20\\
    \end{tabular}\label{tab:BiSe_L_eps}
    \caption{Lorentz model parameters for \BiSe.}
    
    \label{tab:Lorentz_BiSe}
\end{table}

As discussed in the main text, the experimental data show additional peaks at $\omega\sim1600$ cm$^{-1}$ and $\omega\sim1800$ cm$^{-1}$. These peaks can be included in our MLFDM calculations (see Fig. \ref{fig:ISS}) by assuming a multilayer description of the \BiSe $ $ nanoribbon. We define an additional thin surface layer (of 5 nm) on top of the thick bulk layer (on 80 nm). In particular, the dielectric responses of the two layers of the TINR are modelled with a bare Drude model (Eq. \ref{eq:Drude}) for the bulk layer and two additional Lorentz terms (Eq. \ref{eq:Drude}, Tab.\ref{tab:Lorentz_BiSe}) for the thinner layer which captures the measured response well. We note that introducing this additional layer on the bottom of the TINR yields an identical response, except for a lower amplitude, meaning we are unable using the model to determine which, if not both, surfaces this response originates from.



\section{Frequency spectra}
In this section, we extend the analysis of the frequency-dependent spectra of the near-field signal shown in Fig. 3 in the main text. In particular, we provide further details on the post-processing required to obtain the frequency spectra from the two sets of measurements in analysis (approach curves and 2D scans), and the modelling of the carrier concentration of the \BiSe $ $ nanoribbon atop different substrates. 
\begin{figure}[!t]
    \centering
    \includegraphics[width=1\columnwidth]{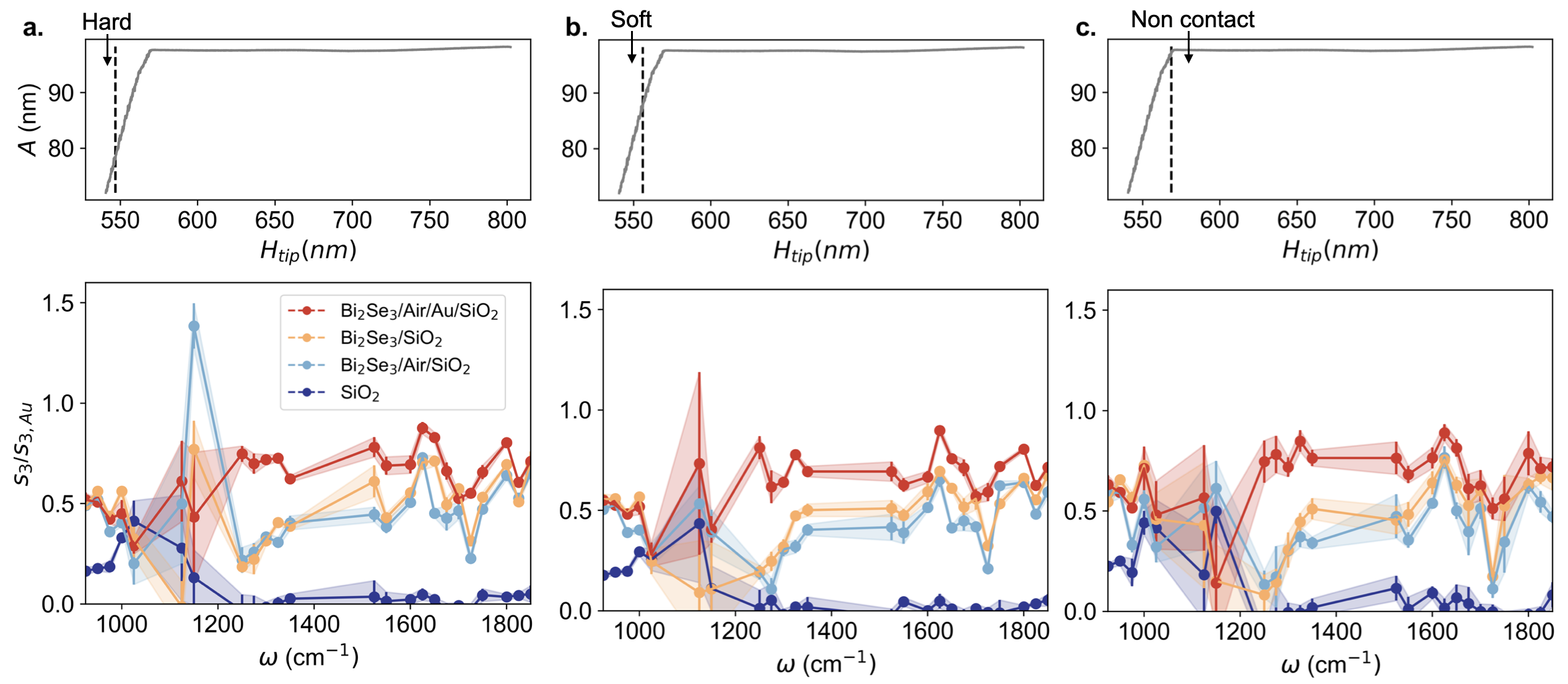}
    \caption{\textbf{Post-processing of approach curve data for frequency spectra.} Frequency dependent behaviour of the near-field signal measured on different positions on the sample. The measurements shown in the panels are extracted from the approach curves at fixed position of the tip respectively in hard, soft and non contact tapping mode. All the signals are renormalised with respect to Au. \textbf{a.} Second ($s_{2}$), \textbf{b.} third ($s_{3}$) and \textbf{c.} fourth ($s_{4}$) harmonics of the near-field signal.}
    \label{fig:eta_omega_dh_cut}
\end{figure}
\begin{figure}[t]
    \centering
    \includegraphics[width=1\columnwidth]{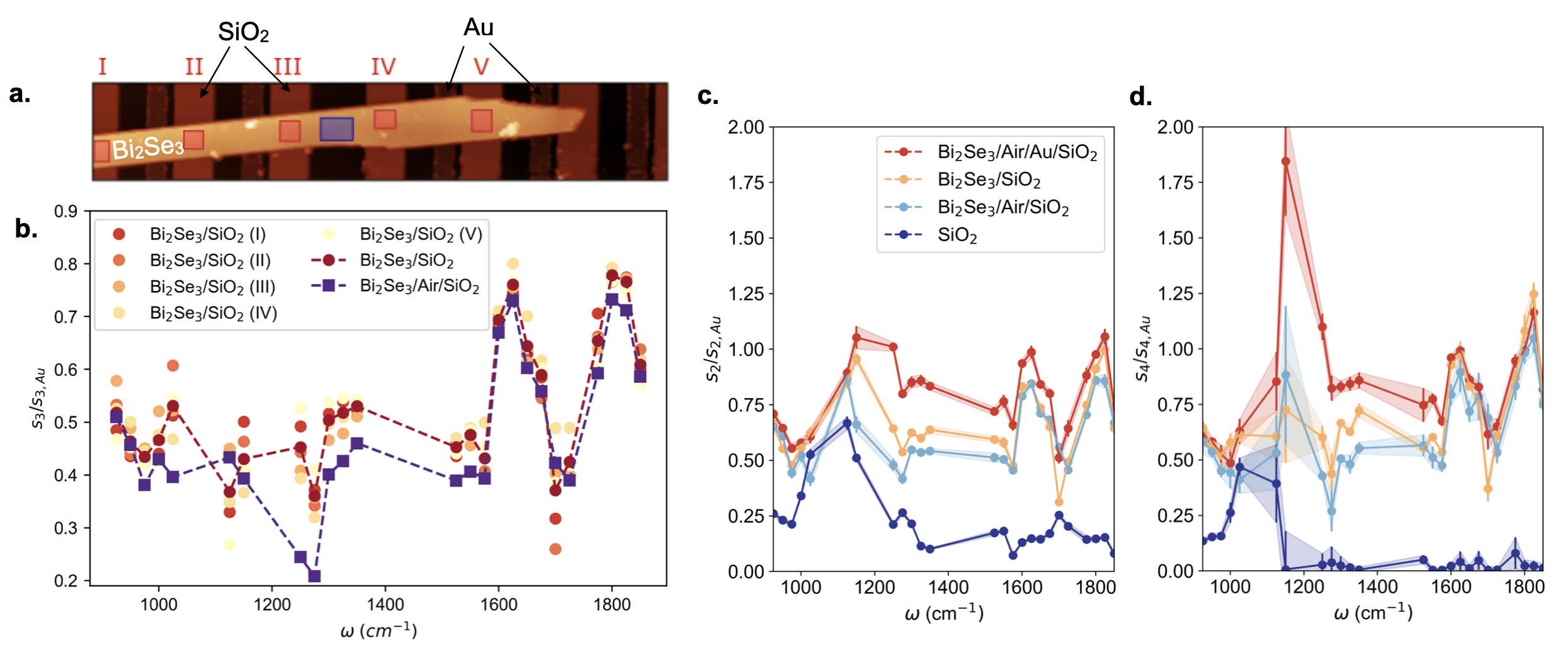}
    \caption{\textbf{Post-processing of surface scans for  frequency spectra.} \textbf{a.} AFM topography with different regions highlighted. The vertical size of the scan is 2.5 $\mu$m.  The $s_3$ response averaged across these different regions and as a function of IR wavelength is shown in \textbf{b.} together with the average of all regions I-V. The signal from the \BiSe/Air/\SiO $ $ region is consistently lower than all the other regions where the \BiSe $ $ rests directly on \SiO. Frequency dependent behaviour of the second ($s_2$) \textbf{c.} and fourth ($s_4$) \textbf{d.} harmonics of the near-field signal. All the signals are re-normalised with respect to Au.}
    \label{fig:eta_omega_2D_scans}
\end{figure}

\begin{figure}[!t]
    \centering
    \includegraphics[width=1\columnwidth]{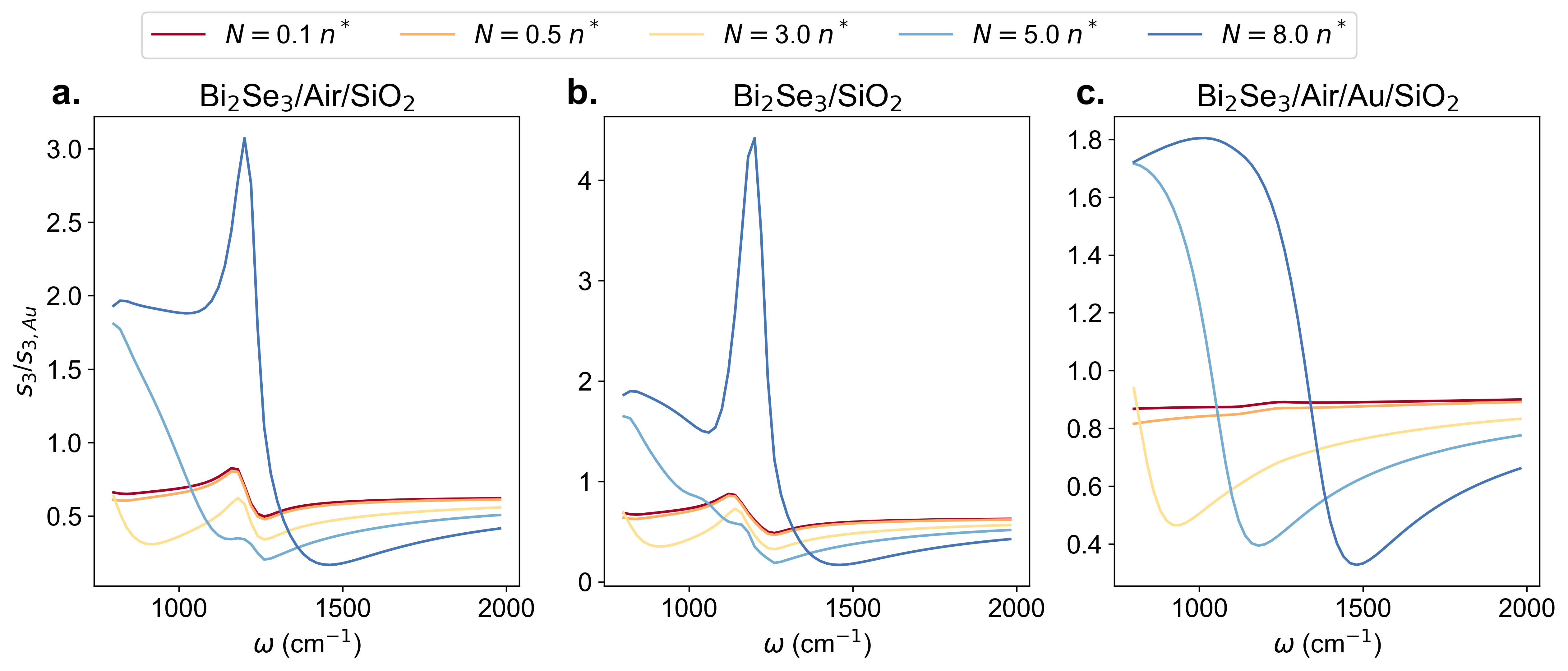}
    \caption{\textbf{Tuning carrier concentration values using the MLFDM.}  $s_3$ (normalised to Au) computed with the MLFDM for the \BiSe $ $ nanoribbon on three different substrates: \textbf{a.} \BiSe/Air/\SiO, \textbf{b.} \BiSe/\SiO $ $ and \textbf{c.} \BiSe/Air/Au/\SiO. For each stack, different values of carrier concentrations were considered. The tip parameters are fixed to $R=50$ nm, $A=75$ nm, $L=600$ nm and $g=0.97$. $n^*=10^{19}$ cm$^{-3}$.}
    \label{fig:MLFDM_eta_omega_N_car_test}
\end{figure}

\paragraph{Spectra from approach curves.}
The frequency-dependent values of the near-field signal can be extracted from the post-processing of the approach curves. As already outlined in section 3, the approach curves are recorded with the tip fixed atop different sample locations and by varying its distance from the sample. The locations are chosen such as to have sufficient separation between regions in order to avoid  finite-size effects.  In Fig. \ref{fig:eta_omega_dh_cut} we report the near-field values extracted at a fixed distance of the tip for different frequency values. Notice that the three columns represent the measurement extracted at the three different tapping regimes (hard, soft and non contact mode) as indicated in Fig. 2b of the main text. 
 
The major response is measured when the \BiSe $ $ nanoribbon is on top of Au, and the relative position of the yellow and blue line emphasises the different response of the \BiSe $ $ when it is suspended or in contact with the \SiO $ $ substrate, and we consistently observe a difference in the range 1200-1600 cm$^{-1}$. 
Inconsistent features across the three panels are observed only between 1000-1200 cm$^{-1}$. In this frequency range the sSNOM signal is generally very weak which results in increased noise and uncertainty.

\paragraph{Spectra from surface scans.}
In contrast to approach curves, the measurements extracted from the 2D surface scans are for a fixed distance of the tip from the sample, and scans cannot be taken in what we here define as the non contact sSNOM regime where the FDM model is valid. For all scans we use an AFM amplitude set-point of $\approx 90$\% of the free oscillation amplitude, which is within the \textit{soft} tapping regime.

To validate the observation that the optical response of the \BiSe $ $ nanorribbon differs whether suspended or in contact with the \SiO  $ $ substrate, we take into consideration measurements of the \BiSe/\SiO $ $ performed on different positions along the \BiSe, as shown in Fig. \ref{fig:eta_omega_2D_scans}a, chosen such as to have maximal separation and avoiding significant contributions due to any finite-size effects (here relevant for length-scales $\lesssim 350$ nm).  
Again we observe in Fig. \ref{fig:eta_omega_2D_scans}b that the signal measured on the stack \BiSe/\SiO $ $ is larger than the signal measured in presence of the air gap (\BiSe/Air/\SiO), for all the different areas analysed. 

In Fig. \ref{fig:eta_omega_2D_scans}c some weak signatures of the two peaks at 1600 and 1800 cm$^{-1}$ can be seen even when the tip is atop \SiO $ $ alone, due to far-field coupling of the lower harmonics. However, for higher harmonics in Fig. \ref{fig:eta_omega_2D_scans}d these peaks disappear, indicating that for higher harmonics finite-size effects are irrelevant on length-scales $\gtrsim 0.4$ $\mu$m, the distance away from the \BiSe $ $ that the bare \SiO $ $ was analysed. This also confirms that the two additional peaks at 1600 and 1800 cm$^{-1}$ indeed originate from the \BiSe $ $ itself.

\paragraph{MLFDM results.}
In Fig. \ref{fig:MLFDM_eta_omega_N_car_test} we report the results obtained for different \BiSe $ $ carrier concentrations. The theoretical model allowed us to establish a finite range of possible carrier concentrations which properly fit the frequency dependent behaviour. The comparison of the different panels reveal that the system analysed is correctly described only if the carrier concentration is in the range $3-4 \cdot10^{19}$ cm$^{-3}$.

\end{document}